\documentclass[aps, prl, reprint, twocolumn, 
amsmath,amssymb,superscriptaddress,shortbibliography]{revtex4-2}
\usepackage[version=4]{mhchem}
\usepackage[utf8]{inputenc}
\usepackage[T1]{fontenc}
\usepackage{graphicx,graphics,epsfig}
\usepackage{amsmath, amssymb,amsfonts} 
\usepackage{bm,times,xspace,mhchem}  
\usepackage{mathrsfs}
\usepackage{color}
\usepackage[colorlinks, citecolor=blue, linkcolor=red, citecolor=blue, urlcolor=blue]{hyperref}
\usepackage{dcolumn}
\usepackage{float}
\usepackage{braket}
\usepackage{verbatim}
\usepackage{soul}
\usepackage {lineno}
\usepackage{xr}
\usepackage{hyperref}

\begin{document}

\title{Complexity Powered Machine Intelligent Classification of Quantum Many-Body Dynamics}


\author{Zhaoran Feng}
\affiliation{Center for Phononics and Thermal Energy Science, China-EU Joint Lab on Nanophononics, Shanghai Key Laboratory of Special Artificial Microstructure Materials and Technology, School of Physics Science and Engineering, Tongji University, Shanghai 200092, China} 

\author{Jiangzhi Chen}%
\affiliation{Center for Phononics and Thermal Energy Science, China-EU Joint Lab on Nanophononics, Shanghai Key Laboratory of Special Artificial Microstructure Materials and Technology, School of Physics Science and Engineering, Tongji University, Shanghai 200092, China}

\author{Ce Wang}
\email{phywangce@gmail.com}
\affiliation{Center for Phononics and Thermal Energy Science, China-EU Joint Lab on Nanophononics, Shanghai Key Laboratory of Special Artificial Microstructure Materials and Technology, School of Physics Science and Engineering, Tongji University, Shanghai 200092, China}

\author{Jie Ren}
\email{xonics@tongji.edu.cn}
\affiliation{Center for Phononics and Thermal Energy Science, China-EU Joint Lab on Nanophononics, Shanghai Key Laboratory of Special Artificial Microstructure Materials and Technology, School of Physics Science and Engineering, Tongji University, Shanghai 200092, China} 
\affiliation{Shanghai Research Institute for Intelligent Autonomous Systems, Tongji University, Shanghai 200092, P. R. China}%

\date{\today}

\begin{abstract}
Identifying and classifying quantum phases from measurable time series in many-body dynamics have significant values, yet lack details and face formidable challenges, requiring profound knowledge of physicists. 
Here, to achieve a purely data-driven machine intelligent classification, we introduce a temporal fluctuation-amplified distance measure that captures the inherent temporal fluctuation complexity  of dynamic evolution series in different quantum many-body phases. Significantly, the introduction of complexity-powered distance leads to remarkable improvements of unsupervised manifold learning of quantum many-body dynamics, as exemplified in models such as the discrete time crystal, Aubry-Andr\'e, Quantum East , and Feingold-Peres models. Our method does not require any prior knowledge and exhibits effectiveness even in imperfect, disordered, and noisy situations that are challenging for human scientists.
Successful classification of dynamic phases in many-body systems holds the potential to enable crucial applications, including identification of tsunamis, earthquakes, catastrophes and future trends in finance.
\end{abstract}

\maketitle

\textit{Introduction}-
Non-equilibrium quantum many-body systems have garnered significant research interest, \textit{spanning} active fields such as Floquet evolution \cite{Marin2015Floquet_advances,rudner2020band,RevModPhys.89.011004}, quench dynamics \cite{RevModPhys.83.863,Luca2016AP}, and open quantum systems \cite{RevModPhys.88.021002,RevModPhys.89.015001,RevModPhys.93.015008}. These dynamical processes give rise to remarkable phenomena including discrete time crystals (DTC) \cite{wilczek2012quantum_prl,zhang2017observation_nature,yao2017discrete_prl,RevModPhys.95.031001,Chen2023_nc}, many-body localization (MBL) \cite{RevModPhys.91.021001,RevModPhys.80.885,Guo2021_np}, and prethermalization \cite{ueda2020quantum_nrp,le2023observation_nature}. Nevertheless, theoretical characterization of such systems faces substantial challenges:~both analytical methods and numerical simulations become intractable with increasing system size. Even in non-interacting systems, phase classification grows remarkably complex, as exemplified by the 54 non-equivalent topological classes identified in open systems~\cite{Gong2018Topological_prx}. To address these limitations, we investigate hidden dynamical signatures and develop a purely data-driven framework leveraging artificial intelligence for quantum many-body phase classification. 

Recent advances have revealed machine learning's potential in addressing condensed matter physics, quantum domains, and topological physics \cite{RevModPhys.91.045002,rem2019identifying,kaming2021unsupervised}. For those systems that are not well-understood by human scientists, unsupervised machine learning offers a valuable and effective method for identifying and classifying patterns in data without the need for prior knowledge~\cite{van2017_np,Lu2020Extracting_prx,WangLei_2016_prb,Hu2017_pre,Wang2017Machine_prb,Rodriguez-Nieva2019_np,YangLong2020_prl,Che2020Topological_prb,Yu_LiWei_2021_prl}. Significant progress has been achieved through techniques ranging from principal component analysis applied to classical statistical models \cite{WangLei_2016_prb,Hu2017_pre,Wang2017Machine_prb} to diffusion mapping for topological transitions \cite{Rodriguez-Nieva2019_np,YangLong2020_prl,Che2020Topological_prb,Yu_LiWei_2021_prl}. However, existing approaches predominantly utilize static system features—such as microscopic configurations or ground-state wave functions—neglecting critical temporal information embedded in dynamical processes. This limitation motivates our search for new paradigms capable of processing time-series data with inherent temporal correlations.

\begin{figure*}
    \centering
    \includegraphics[width=0.85\linewidth]{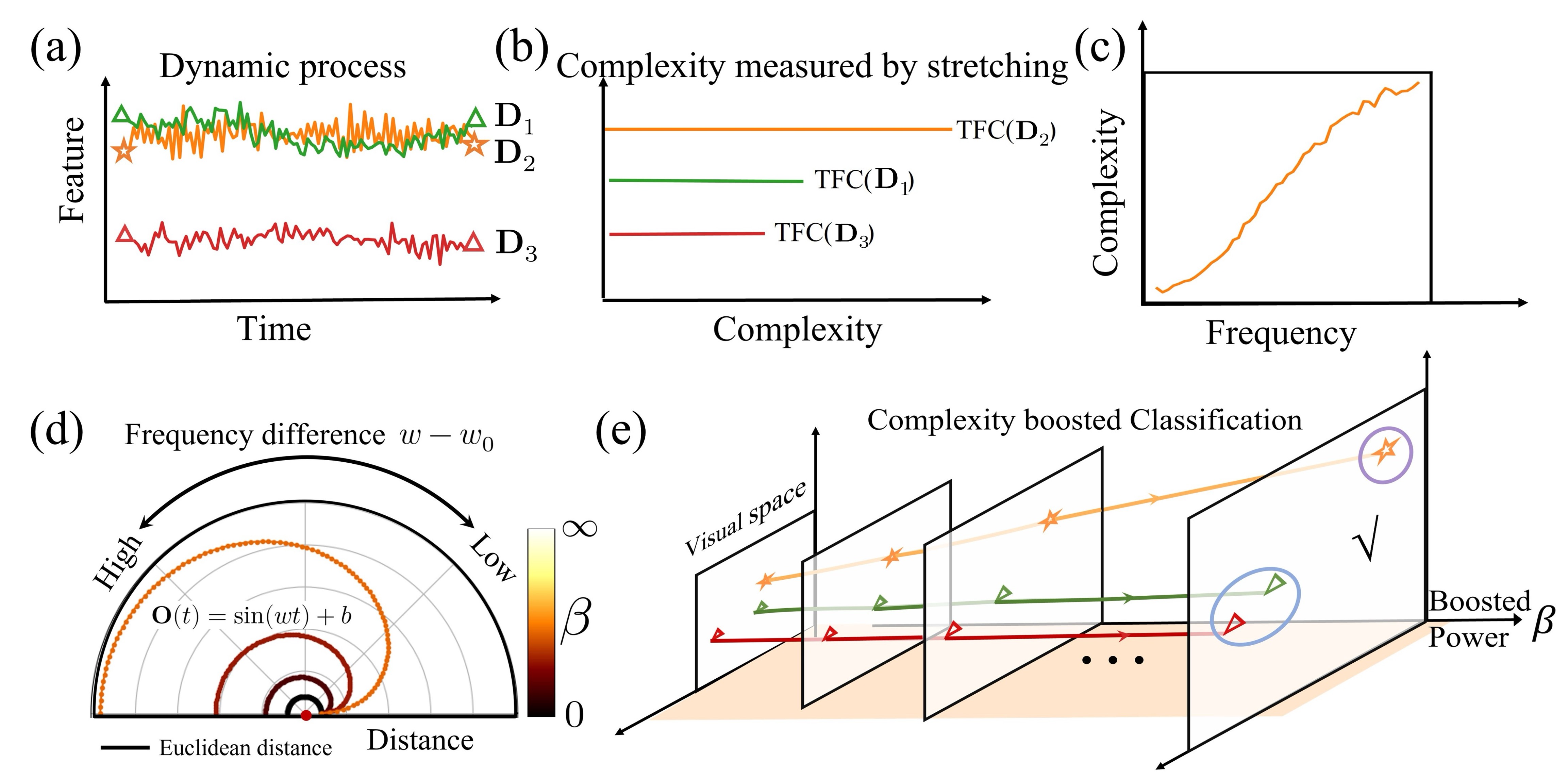}
    \caption{\textbf{Schematic framework of complexity-powered quantum dynamics analysis.} 
    \label{schematic_diagram}
    \textbf{a}, Temporal evolution of quantum many-body dynamical processes;  \textbf{b}, TFC quantification via configuration-space stretching metrics; \textbf{c}, Frequency-dependent complexity scaling: TFC increases with the dominance of high-frequency components; \textbf{d}, Frequency amplification mechanism: complexity enhances inter-series frequency discrimination, while power boosting exponentially magnifies these distinctions;  \textbf{e}, Classification performance based on the TFCAD framework demonstrates progressive improvement as $\beta$ parameters are augmented (see Supplemental Material [SM Sec.1]~\cite{SM}).}
\end{figure*}

In this Letter, we introduce the temporal fluctuation complexity (TFC) \cite{Gustavo2011_book,RevModPhys.86.1169} metric, which quantifies dynamic features through time-series analysis. Building on this foundation, we propose the Temporal Fluctuation Complexity Amplified Distance (TFCAD)—a distance operator that measure the similarity between the time-varying dynamic observables.

When combined with diffusion mapping, the TFCAD-powered diffusion framework achieves dual breakthroughs: precise classification of quantum many-body phases and construction of new phase diagrams, consistently outperforming Euclidean-based approaches.

\textit{Complexity enhancement}-
We first define the general form of time series data as $\mathbf{D} \equiv [\mathbf{O}(t_0),\mathbf{O}(t_0+\delta t),...,\mathbf{O}(t_h)]$. This represents a set of values for one or more physical quantities at equally spaced time intervals. The halting time  $t_h = t_0 + (L_s - 1)\delta t $ defines a length-$L_s$ sequence, where $\mathbf{O}(t_0 + (l - 1)\delta t)$ is abbreviated as $\mathbf{O}_l$. Data $\mathbf{D}_{i}$ are indexed by $i$ to denote distinct system parameters. 

Defining the inter-dataset distance metric constitutes a crucial prerequisite when employing unsupervised learning methods to cluster data and identify phase transitions. One natural choice is to use the Euclidean distance as $M_{ij}= \lVert \mathbf{D}_i - \mathbf{D}_j \rVert_2$ which is widely used in previous papers \cite{Durr2019_prb,RodriguezNieva2019_np,Zvyagintseva2021_commphys}. However, the similarity between two sequences is determined not only by their distance at corresponding time, but also by their fluctuation patterns. As shown in  Fig.~\ref{schematic_diagram}(a), two sequences with a small Euclidean distance might have very different fluctuation patterns. Conversely, two sequences displaying similar patterns may have a large Euclidean distance. Therefore, we need to propose a distance definition that balances the Euclidean distance and the similarity of fluctuation patterns between sequences.

The TFC measures cumulative variations in temporal sequences through trajectory geometry~\cite{Gustavo2011_book}.  For a discrete sequence $\mathbf{D}$ with state vectors $\mathbf{O}_l \in R^{m} $, the TFC is defined as: 
\begin{equation}  
C(\mathbf{D})=\sqrt{\sum_{l=1}^{L_s-1}
\left\lVert \mathbf{O}_l-\mathbf{O}_{l-1}\right\rVert_2^{2}}
\end{equation}
This metric geometrically corresponds to the stretching length of the sequence trajectory, as demonstrated in Fig.~\ref{schematic_diagram}(b). 
Fig.~\ref{schematic_diagram}(c) illustrates that TFC scales with the dominant oscillatory frequency of the temporal patterns. It effectively captures high-frequency features, where rapid state transitions lead to more pronounced trajectory stretching.

We apply TFC to construct distances between sequences. The TFCAD is given by the following formula~\cite{ij}: 
\begin{equation} 
\tilde{M}_{ij} = E^{\beta}_{ij} M_{ij} 
\label{eq:1} 
\end{equation}
Here $M_{ij} $ denotes the Euclidean distance between elements $i$ and $j$, the complexity-boosted factor $E_{ij} = \text{max} \{C_i+\delta,C_j+\delta\}/\text{min}\{C_i+\delta,C_j+\delta\} $ with $\delta\xrightarrow{}0^+$ as an infinitesimal constant to ensure numerical stability (preventing zero denominators), and $\beta\ge 0$ is the boosted power to adjust the weight of the complexity enhancement. 

For a non-zero $\beta$, when two sequences $\mathbf{D}_i,\mathbf{D}_j$ have similar TFC, $E^{\beta}_{ij}$ is close to unitary, meaning that their distance $\tilde{M}_{ij}$ is approximately the original Euclidean distance $M_{ij}$. However, when the TFC of the two sequences differ greatly, $E^{\beta}_{ij}$ will be much larger than one, significantly increasing  the distance $\tilde{M}_{ij}$ compared to $M_{ij}$.
To demonstrate TFCAD's discriminative capability, we analyze a toy dataset where each sequence follows $\mathbf{O}(t) = \sin(wt) + b $ with fixed offset $b$ and varying frequency $w$. 
Fig.~\ref{schematic_diagram}(d) displays TFCAD  between these sequences and a reference $\mathbf{O}_{r}(t) = \sin(w_0 t)$ under varying $\beta$. 
While sequences exhibit identical Euclidean distance ($\beta=0$), non-zero $\beta$ values reveal frequency-dependent separation:  TFCAD increase  with frequency disparity $\Delta  w=w-w_0 ( w_0<w )$, amplifying significantly for larger $\beta$. 

For the three sequences in Fig.~\ref{schematic_diagram}(a), we observe  $C(\mathbf{D_2})> C(\mathbf{D_1})\approx C(\mathbf{D_3})$ reflecting the pronounced spectral disparity in high-frequency amplitudes between $\mathbf{D_2}$ and $\mathbf{D_1}/\mathbf{D_3}$. In Fig.~\ref{schematic_diagram}(e), $\mathbf{D_1}$ and  $\mathbf{D_2}$ are closer under Euclidean distance. 
As  $\beta$ increases from zero, $\mathbf{D_1}$ diverges from $\mathbf{D_2}$ and converges toward $\mathbf{D_3}$, showing that TFCAD enables significantly different classification at large $\beta$. 

\begin{figure*}
\centering
    
\includegraphics[width=0.9
\linewidth]{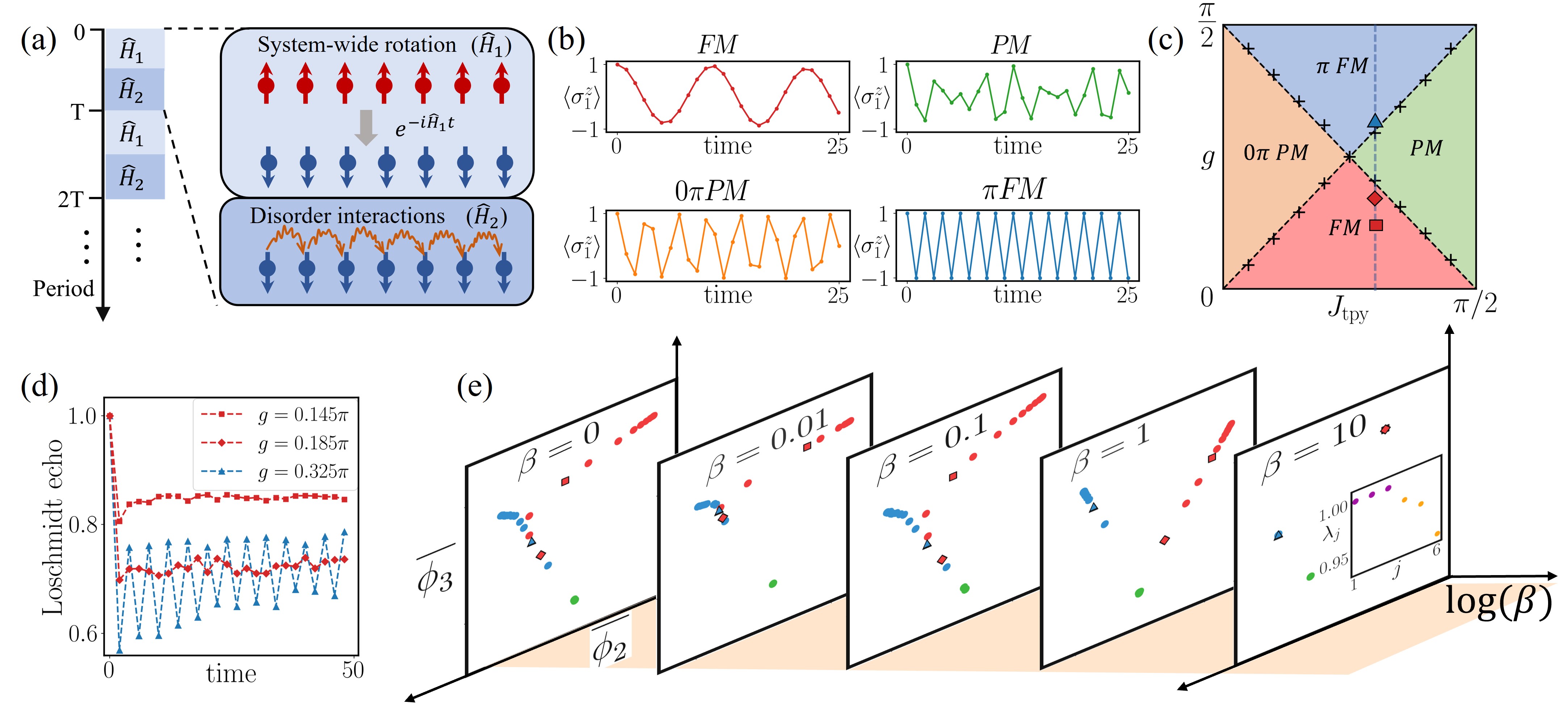}
    \caption{\textbf{The unsupervised phase classification of DTC model}. \textbf{a}, Floquet dynamics of DTC under dual-pulse driving: global rotation ($\hat{H}_1$) followed by disorder-enhanced nearest-neighbor coupling ($\hat{H}_2$);  \textbf{b}, Site-1 Z-operator expectation value evolution;  \textbf{c}, The two-dimensional phase diagram is generated through the application of the TFCAD framework to dynamical sequences acquired under systematically varied driving field strengths $g$ and interaction parameters $J_{typ}$. The cluster boundaries marked by the black 
    crosses is consistent with the theoretical phase diagram (the orange / red / green / blue region represents the  $0\pi$  paramagnetic / ferromagnetic /  paramagnetic / $\pi$ ferromagnetic phase); \textbf{d}, The LE data collected at $J_{\text{typ}}=0.3\pi$  for $g=0.145\pi$ (red square), $g=0.185\pi$ (red diamond), and $g=0.325\pi$ (blue triangle) in panel (c) correspond to the three easily confusable cases in the $\beta=0$ regime; \textbf{e}, The classification results are derived from data represented by the blue dotted line in panel (c) (corresponding to varying g at fixed $J_\text{typ} = 0.3\pi$), which were used to generate the categorical mapping. The low-dimensional representation is constructed through an orthogonal transformation of the transfer matrix's second and third eigenstates: 
 $\bar{\phi}_{2} = \cos\alpha\phi_2-\sin\alpha\phi_3$ and $\bar{\phi}_{3}= \sin\alpha\phi_2+\cos\alpha\phi_3$, where the parameter $\log(\beta)$ spans  $-\infty$ to $2$.  The inset in panel (e) displays the first six eigenvalues of the transfer matrix.}
    
    \label{dtc_withcf}
\end{figure*}

\textit{Diffusion map}-
Once the distance between the data is defined, we can discover the underlying manifold for a given dataset by the diffusion map algorithm~\cite{lidiak2020unsupervised}, thereby achieving dimension reduction and clustering. The similarity matrix is then given by a Gaussian kernel, 
\begin{equation}
    K_{ij}={\rm exp}
    \left(-\frac{\tilde{M}_{ij}}{2\epsilon L_s^2}\right)
\label{similarity1}
\end{equation}
Here $\epsilon$ is the adjustable Gaussian kernel coefficient. Using Eq.~(\ref{similarity1}), we construct the similarity matrix for  $N$ randomly sampled Hamiltonians in parameter space, where $K_{ij}\xrightarrow{}1$ indicates similarity between two samples and $K_{ij}\xrightarrow{}0$ otherwise. The probability transition matrix is defined as $P_{ij}=K_{ij}/\sum_{j=1}^{N}K_{ij}$ to describe the progress of diffusion. After $t$ steps, the diffusion distance between sample $i$ and $j$ on the manifold is $d^{t}_{ij}=\sum_{n}[(P_{in}^{t}-P_{jn}^{t})^2/\sum_{n'}K_{nn'}]=\sum_{n=1}^{N-1}\lambda_{n}^{2t}[(\phi_{n})_i-(\phi_{n})_j]^2$, where $\phi_n$ denotes the $n$-th right eigenvector of $\hat{\mathbf{P}}$ and $\lambda_{n}$ is the $n$-th eigenvalue ($n=0,1,...,N-1$). After long diffusion time, the first few components $\phi_n$ that have the largest eigenvalues $\lambda_n\approx1$ will dominate in the manifold diffusion process. Consequently, the essential information of the manifold diffusion distance is contained within these few components, and they form a low-dimensional representation of the original sequential data. Phase classification can then be derived by clustering these compact representations. Next, we will apply this diffusion map algorithm based on TFCAD to different dynamical quantum systems. 

\begin{figure*}[tp]
\centering
\includegraphics[width=0.9\linewidth]{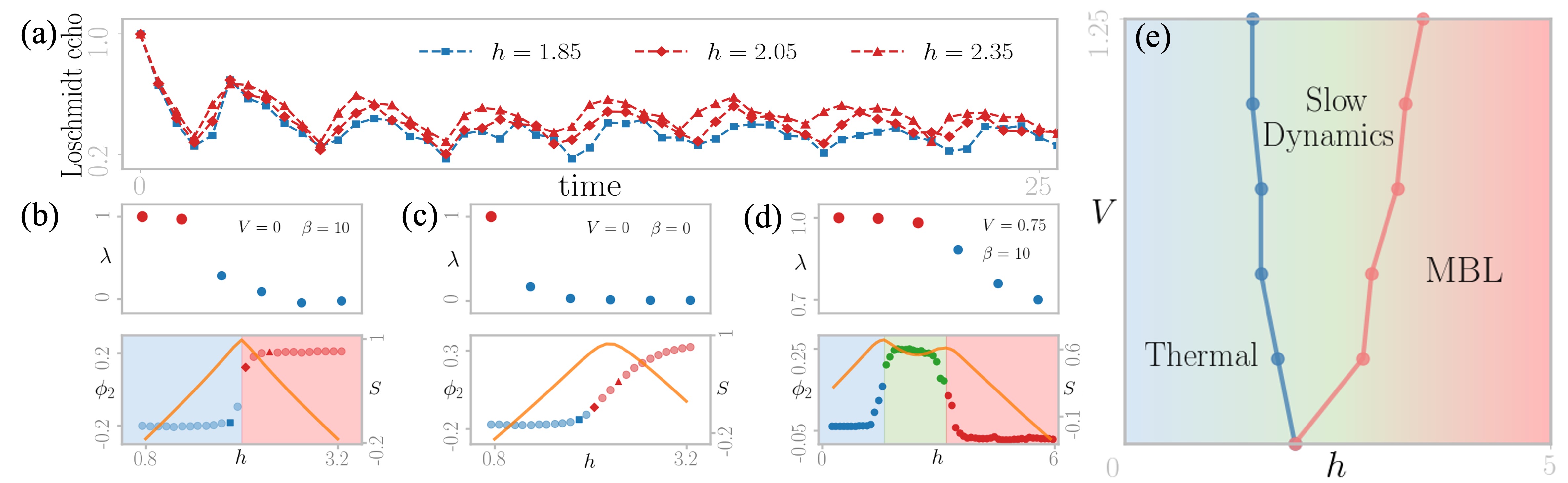}
\caption{\textbf{The unsupervised clustering of AA model.} \textbf{a}, LE data for $V=0$ at three disorder strengths: $h=1.85$ (blue squares), $h=2.05$ (red diamonds), and  $h=2.35$ (red triangles);
\textbf{b}, First six eigenvalues of transfer matrix $\hat{\mathbf{P}}$  computed using Euclidean distance (upper panel) and TFCAD (lower panel); \textbf{c}, Low-dimensional representation constructed from the second eigenstates of $\hat{\mathbf{P}}$ , classified via Euclidean metric (light shades) and TFCAD (dark shades). $S$ (orange curves) quantify clustering quality for $h$-parameterized configurations; \textbf{d}, First six eigenvalues of $\hat{\mathbf{P}}$  using TFCAD (upper panel); low-dimensional representation derived from the third eigenstates (TFCAD-classified, color-coded) with corresponding $S$ (orange curves);  \textbf{e}, Two-dimensional phase diagram reconstructed from time-dependent density matrix renormalization group (t-DMRG) simulations, with classifications performed using the TFCAD framework.  }
    \label{AA_withcf}
\end{figure*}

\textit{Discrete time crystal model}-
The spontaneous breaking of time-translation symmetry is a dynamical phenomenon that exclusively arises  in non-equilibrium systems~\cite{ponte2015manybody_prl,lazarides2015fate_prl,khemani2016phase_prl}. In order to realize the breaking of time symmetry, periodically driven Floquet systems are widely concerned~\cite{wilczek2012quantum_prl,zhang2017observation_nature,RevModPhys.95.031001,yao2017discrete_prl,xiaopeng2023subexponential_prb,Chen2023_nc}. In these systems, discrete time-translation symmetry may be further broken, manifested as double period oscillations for some physical quantities. We investigate the Floquet quantum dynamics of a one-dimensional disordered spin chain governed by the following Floquet Hamiltonian~\cite{yao2017discrete_prl}:

\begin{equation} \hat{H}_{\text{DTC}}(t)=\begin{cases}\hat{H}_1 =  g\sum_{j=1}^L\hat{\sigma}_j^x,\quad&0\leq t\leq t_1\\ \hat{H}_2 =  \sum_{j=1}^{L-1}J_j\hat{\sigma}_j^z\hat{\sigma}_{j+1}^z.\quad&t_1\leq t\leq t_1 + t_2\end{cases}
\end{equation}
Here, $\hat{\sigma}_j^{\gamma} (\gamma=x,y,z)$ denotes  the Pauli operator acting on the $j^{\text{th}}$ spin, $g$ represents  the Rabi frequency, and $J_j$ is the random coupling strength between spins $j$ and $j+1$. The distribution of $J_j$ is described by a typical strength $J_{\text{typ}}$ as $\ln J_{\text{typ}} = \frac{1}{L} \ln J_j$, where their logarithms have a standard deviation $\sigma_{J} = 0.02\pi$. We choose $t_1 = t_2 = 1$ such that the Floquet driven has a period of $T=2$. The so-called \textit{discrete time symmetry broken} can be realized by setting $g = \frac{\pi}{2}$, where a product state with all the spins parallel to $z$-direction will return back to itself after two Floquet periods. Interestingly, under the random interacting process captured by $\hat{H}_2$, this double period temporal pattern is stable even when $g$ is deviated from $\pi/2$.  

Fig.~\ref{dtc_withcf}(a) depicts the Floquet evolution of a spin chain, where two Hamiltonian are sequentially applied over time: a nearly $\pi$ global spin flip ($\hat{H}_1$) followed by disorder interactions ($\hat{H}_2$). 
The varying time-evolved magnetization of spins  $\left\langle\hat{\sigma}_i^{z}(t)\right\rangle$ exhibit different sub-harmonic responses of the system to the Floquet Hamiltonian  in Fig.~\ref{dtc_withcf}(b).
A complete phase diagram for this Floquet process contains four different phases as shown in Fig.~\ref{dtc_withcf}(c)~\cite{Moessner2017_np}. 

Here, we employ the Loschmidt echo (LE) as the temporal feature~\cite{Quan_H_T_2006_prl,GORIN2006_pr,yan2020information_prl}. It has been established as an experimentally accessible observable in quantum many-body simulators~\cite{martinez2016real,jurcevic2017direct,singh2019quantifying,karch2025probing}, thereby reinforcing its practical relevance.
The LE is defined by the correlation between the initial state $|\psi_0\rangle $ and the state $|\psi_t \rangle $ at time $t$ as $\mathcal{L}(t) = |\langle \psi_0 | \psi_t \rangle |^2$. Note that this model is solvable under Jordan-Wigner transformation, allowing exact computation of the LE for sufficiently large systems.  The input sequential data is then collected at each Floquet period as $\mathbf{O}_l = \mathcal{L}(2l)^{1/L}$ where the exponential factor $1/L$ is introduced to eliminate the effect of the Anderson's orthogonal catastrophe. The dataset comprises LE time series  gathered at varying values of $g$ under a fixed $J_\text{typ}$.

Application of the TFCAD-based diffusion map reveals three dominant eigenvalues ($\lambda_n \approx 1$) in the spectral decomposition, corresponding to three distinct phases. We determine the phase transition points by combining the "learning by confusion" framework~\cite{van2017_np} with silhouette coefficient ($S$) clustering. The $S$ metric, which evaluates clustering quality based on intra-cluster cohesion and inter-cluster separation, is analyzed in detail in [SM Sec.~3]~\cite{SM}, and its use for determining phase-transition points (including the learning-by-confusion procedure) is described in [SM Sec.~4]~\cite{SM}. Systematic variation of $J_\text{typ}$ yields a phase diagram in quantitative agreement with theoretical predictions as shown in Fig.~\ref{dtc_withcf}(c). 

To highlight the critical role of TFCAD, we compare our method with diffusion maps employing increasing boosted power $\beta$ for the $J_\text{typ} = 0.3\pi$ dataset. As shown in Fig.~\ref{dtc_withcf}(e), the $\beta=0$ case fails to distinguish those three phases, while increasing $\beta$  gradually identifies all three phases correctly. 
Three representative points from Fig.~\ref{dtc_withcf}(c) are analyzed: a $\pi$-ferromagnetic phase marker (blue triangle, $g=0.325\pi$) and two ferromagnetic phase markers (red diamond/square, $g=0.185\pi/0.145\pi $). Their LE sequences (Fig.~\ref{dtc_withcf}(d)) reveal that Euclidean distance ($\beta=0$) erroneously groups the red diamond with the blue triangle. Increasing $\beta$ amplifies the TFCAD between these states, achieving perfect cluster separation at $\beta=10$ (Fig.~\ref{dtc_withcf}(e)). Extended analysis of the DTC model are detailed in [SM Sec.~2]~\cite{SM}. 

\textit{Aubry-Andr\'e model}-
To further demonstrate the universality of our method, we consider the dynamic process for the Aubry-Andr\'e (AA) model~\cite{Aubry_1980}. The AA model, originally formulated for non-interacting particles in quasi-periodic potentials, is here generalized through explicit electron-electron interactions. This extension yields the modified Hamiltonian:
\begin{equation}
    \hat{H}_{AA}=\sum_{j=0}^{L-1}\left[h_j\hat{n}_j+J(\hat{c}_j^\dagger\hat{c}_{j+1}+\hat{c}_{j+1}^\dagger\hat{c}_j)+V\hat{n}_j\hat{n}_{j+1}\right]
\end{equation}
Here, $\hat{c}_j(\hat{c}_j^{\dagger})$ represents the fermion annihilation (creation) operator on site $j$, and $\hat{n}_j=\hat{c}_j^{\dagger} \hat{c}_j$ denotes the corresponding density operator. The quasi-periodic onsite potential is denoted by $h_j = h\cos(2\pi kj+\delta)$, where $k=\frac{\sqrt{5}-1}{2}$. The terms $J$ is the nearest-neighbor hopping and set to be $1$ as the energy unit, while  $V$ represent the nearest-interaction. In the non-interacting limit ($V=0$), the system exhibits a single-particle localization-delocalization transition at the self-dual critical point $h = 2$.  
In the interacting case ($V>0$), a thermal-MBL phase transition also emerges, though its critical parameters remain unresolved in current studies~\cite{Iyer2013_prb,xu2019butterfly_prr}.

We analyze the phase transition characteristics in the AA model using our method. The dynamical evolution is initiated from a fixed product state  $| \psi_0 \rangle$ under $\hat{H}_{AA}$, specifically constructed as a $1/4$-filled configuration with particles uniformly distributed at sites $j = 0 \pmod{4}$. Input sequential data consists of LE values $\mathbf{O}_l = \mathcal{L}(t)^{1/N_p}$ sampled at discrete times $t =(l-1)\delta t$, where $N_p$ denotes the total particle number. 
We introduce an exponential factor of $1/N_p$ to mitigate the effect of Anderson's Orthogonality Catastrophe. 
The dataset contains LE temporal sequences across different disorder strengths $h$ at fixed interaction $V$.

For the non-interacting case ($V=0$), Fig.~\ref{AA_withcf}(a) displays three LE temporal sequences near the critical point $h=2$, which exhibit visual in-distinguishability across the transition. This observation rationalizes the failure of Euclidean distance-based clustering in pinpointing the phase boundary.  As depicted in Fig.~\ref{AA_withcf}(b), the transition matrix for $\beta=0$ has only one eigenvalue closed to one, indicating the diffusion map cannot distinguish the two phases. However, with $\beta=10$, we find two eigenvalues near to one. Further analysis of the $\phi_2$ components as a function of $h$, integrated with $S$ evaluation (Fig.~\ref{AA_withcf}(c)), uncovers a discontinuous jump at  $h=2$. This  transition aligns exactly with the  theoretical critical value~\cite{Iyer2013_prb}. 

For the interacting case ($V\neq 0$), we performed t-DMRG simulations to generate the LE dynamics. At $V=0.75$, the TFCAD-based diffusion map analysis (Fig.~\ref{AA_withcf}(d)) successfully resolves three distinct phases, in remarkable agreement with the phase classification reported in~\cite{xu2019butterfly_prr}. Within the $\beta=10$ critical scaling regime, we further constructed the full phase diagram (Fig.~\ref{AA_withcf}(e)), which for the first time systematically delineates the thermal phase, slow-dynamics phase, and many-body localized phase — a tripartite classification absent in prior studies. More details of AA model with interactions are presented in [SM Sec.~2]~\cite{SM}.

\textit{Conclusions}-
The results above demonstrate that TFCAD enables a fully data-driven classification of quantum many-body dynamical phases directly from measurable time-series observables. By integrating temporal fluctuation complexity with a nonlinear boosting exponent $\beta$, the framework captures subtle divergences in fluctuation patterns that traditional Euclidean metrics often overlook. When coupled with diffusion mapping, this complexity-enhanced geometry effectively resolves obscured phase boundaries, achieving excellent agreement with established theory across both DTC and AA models.

To ensure the broader reliability and practical utility of this approach, we provide extensive benchmarking and implementation protocols in the Supplemental Material~\cite{SM}. These include systematic validations on the Quantum East [SM Sec.~5], Feingold--Peres [SM Sec.~6], and transverse-field Ising models [SM Sec.~7]. Furthermore, we address critical operational aspects, such as the $S$-saturation criterion for optimizing $\beta$ [SM Sec.~8], the indispensable role of the Euclidean component via ablation studies [SM Sec.~9], and the framework's robustness against finite-size constraints [SM Sec.~10] and sampling-interval variations $\delta_t$ [SM Sec.~11]. Collectively, these findings establish TFCAD as a versatile and interpretable pipeline for dynamical classification, offering potential applications that extend beyond quantum systems to complex time-dependent phenomena in fields ranging from geophysics to finance.

\begin{acknowledgments}
\textit{Acknowledgement}-
We acknowledge the support from  the National Natural Science Foundation of China under Grants No. 12204352 (C.W.), the Natural Science Foundation of Shanghai (No. 23ZR1481200), the Program of Shanghai Academic Research Leader (No. 23XD1423800), and the Opening Project of Shanghai Key Laboratory of Special Artificial Microstructure Materials and Technology.
\end{acknowledgments}

\bibliography{reference}

\begin{thebibliography}{54}%
\makeatletter
\providecommand \@ifxundefined [1]{%
 \@ifx{#1\undefined}
}%
\providecommand \@ifnum [1]{%
 \ifnum #1\expandafter \@firstoftwo
 \else \expandafter \@secondoftwo
 \fi
}%
\providecommand \@ifx [1]{%
 \ifx #1\expandafter \@firstoftwo
 \else \expandafter \@secondoftwo
 \fi
}%
\providecommand \natexlab [1]{#1}%
\providecommand \enquote  [1]{``#1''}%
\providecommand \bibnamefont  [1]{#1}%
\providecommand \bibfnamefont [1]{#1}%
\providecommand \citenamefont [1]{#1}%
\providecommand \href@noop [0]{\@secondoftwo}%
\providecommand \href [0]{\begingroup \@sanitize@url \@href}%
\providecommand \@href[1]{\@@startlink{#1}\@@href}%
\providecommand \@@href[1]{\endgroup#1\@@endlink}%
\providecommand \@sanitize@url [0]{\catcode `\\12\catcode `\$12\catcode
  `\&12\catcode `\#12\catcode `\^12\catcode `\_12\catcode `\%12\relax}%
\providecommand \@@startlink[1]{}%
\providecommand \@@endlink[0]{}%
\providecommand \url  [0]{\begingroup\@sanitize@url \@url }%
\providecommand \@url [1]{\endgroup\@href {#1}{\urlprefix }}%
\providecommand \urlprefix  [0]{URL }%
\providecommand \Eprint [0]{\href }%
\providecommand \doibase [0]{https://doi.org/}%
\providecommand \selectlanguage [0]{\@gobble}%
\providecommand \bibinfo  [0]{\@secondoftwo}%
\providecommand \bibfield  [0]{\@secondoftwo}%
\providecommand \translation [1]{[#1]}%
\providecommand \BibitemOpen [0]{}%
\providecommand \bibitemStop [0]{}%
\providecommand \bibitemNoStop [0]{.\EOS\space}%
\providecommand \EOS [0]{\spacefactor3000\relax}%
\providecommand \BibitemShut  [1]{\csname bibitem#1\endcsname}%
\let\auto@bib@innerbib\@empty
\bibitem [{\citenamefont {Bukov}\ \emph {et~al.}(2015)\citenamefont {Bukov},
  \citenamefont {D'Alessio},\ and\ \citenamefont
  {Polkovnikov}}]{Marin2015Floquet_advances}%
  \BibitemOpen
  \bibfield  {author} {\bibinfo {author} {\bibfnamefont {M.}~\bibnamefont
  {Bukov}}, \bibinfo {author} {\bibfnamefont {L.}~\bibnamefont {D'Alessio}},\
  and\ \bibinfo {author} {\bibfnamefont {A.}~\bibnamefont {Polkovnikov}},\
  }\bibfield  {title} {\bibinfo {title} {Universal high-frequency behavior of
  periodically driven systems: from dynamical stabilization to floquet
  engineering},\ }\href {https://doi.org/10.1080/00018732.2015.1055918}
  {\bibfield  {journal} {\bibinfo  {journal} {Adv. Phys.}\ }\textbf {\bibinfo
  {volume} {64}},\ \bibinfo {pages} {139} (\bibinfo {year} {2015})}\BibitemShut
  {NoStop}%
\bibitem [{\citenamefont {Rudner}\ and\ \citenamefont
  {Lindner}(2020)}]{rudner2020band}%
  \BibitemOpen
  \bibfield  {author} {\bibinfo {author} {\bibfnamefont {M.~S.}\ \bibnamefont
  {Rudner}}\ and\ \bibinfo {author} {\bibfnamefont {N.~H.}\ \bibnamefont
  {Lindner}},\ }\bibfield  {title} {\bibinfo {title} {Band structure
  engineering and non-equilibrium dynamics in floquet topological insulators},\
  }\href {https://doi.org/10.1038/s42254-020-0170-z} {\bibfield  {journal}
  {\bibinfo  {journal} {Nat. Rev. Phys}\ }\textbf {\bibinfo {volume} {2}},\
  \bibinfo {pages} {229} (\bibinfo {year} {2020})}\BibitemShut {NoStop}%
\bibitem [{\citenamefont {Eckardt}(2017)}]{RevModPhys.89.011004}%
  \BibitemOpen
  \bibfield  {author} {\bibinfo {author} {\bibfnamefont {A.}~\bibnamefont
  {Eckardt}},\ }\bibfield  {title} {\bibinfo {title} {Colloquium: Atomic
  quantum gases in periodically driven optical lattices},\ }\href
  {https://doi.org/10.1103/RevModPhys.89.011004} {\bibfield  {journal}
  {\bibinfo  {journal} {Rev. Mod. Phys.}\ }\textbf {\bibinfo {volume} {89}},\
  \bibinfo {pages} {011004} (\bibinfo {year} {2017})}\BibitemShut {NoStop}%
\bibitem [{\citenamefont {Polkovnikov}\ \emph {et~al.}(2011)\citenamefont
  {Polkovnikov}, \citenamefont {Sengupta}, \citenamefont {Silva},\ and\
  \citenamefont {Vengalattore}}]{RevModPhys.83.863}%
  \BibitemOpen
  \bibfield  {author} {\bibinfo {author} {\bibfnamefont {A.}~\bibnamefont
  {Polkovnikov}}, \bibinfo {author} {\bibfnamefont {K.}~\bibnamefont
  {Sengupta}}, \bibinfo {author} {\bibfnamefont {A.}~\bibnamefont {Silva}},\
  and\ \bibinfo {author} {\bibfnamefont {M.}~\bibnamefont {Vengalattore}},\
  }\bibfield  {title} {\bibinfo {title} {Colloquium: Nonequilibrium dynamics of
  closed interacting quantum systems},\ }\href
  {https://doi.org/10.1103/RevModPhys.83.863} {\bibfield  {journal} {\bibinfo
  {journal} {Rev. Mod. Phys.}\ }\textbf {\bibinfo {volume} {83}},\ \bibinfo
  {pages} {863} (\bibinfo {year} {2011})}\BibitemShut {NoStop}%
\bibitem [{\citenamefont {Luca~D'Alessio}\ and\ \citenamefont
  {Rigol}(2016)}]{Luca2016AP}%
  \BibitemOpen
  \bibfield  {author} {\bibinfo {author} {\bibfnamefont {A.~P.}\ \bibnamefont
  {Luca~D'Alessio}, \bibfnamefont {Yariv~Kafri}}\ and\ \bibinfo {author}
  {\bibfnamefont {M.}~\bibnamefont {Rigol}},\ }\bibfield  {title} {\bibinfo
  {title} {From quantum chaos and eigenstate thermalization to statistical
  mechanics and thermodynamics},\ }\href
  {https://doi.org/10.1080/00018732.2016.1198134} {\bibfield  {journal}
  {\bibinfo  {journal} {Adv. Phys.}\ }\textbf {\bibinfo {volume} {65}},\
  \bibinfo {pages} {239} (\bibinfo {year} {2016})}\BibitemShut {NoStop}%
\bibitem [{\citenamefont {Breuer}\ \emph {et~al.}(2016)\citenamefont {Breuer},
  \citenamefont {Laine}, \citenamefont {Piilo},\ and\ \citenamefont
  {Vacchini}}]{RevModPhys.88.021002}%
  \BibitemOpen
  \bibfield  {author} {\bibinfo {author} {\bibfnamefont {H.-P.}\ \bibnamefont
  {Breuer}}, \bibinfo {author} {\bibfnamefont {E.-M.}\ \bibnamefont {Laine}},
  \bibinfo {author} {\bibfnamefont {J.}~\bibnamefont {Piilo}},\ and\ \bibinfo
  {author} {\bibfnamefont {B.}~\bibnamefont {Vacchini}},\ }\bibfield  {title}
  {\bibinfo {title} {Colloquium: Non-markovian dynamics in open quantum
  systems},\ }\href {https://doi.org/10.1103/RevModPhys.88.021002} {\bibfield
  {journal} {\bibinfo  {journal} {Rev. Mod. Phys.}\ }\textbf {\bibinfo {volume}
  {88}},\ \bibinfo {pages} {021002} (\bibinfo {year} {2016})}\BibitemShut
  {NoStop}%
\bibitem [{\citenamefont {de~Vega}\ and\ \citenamefont
  {Alonso}(2017)}]{RevModPhys.89.015001}%
  \BibitemOpen
  \bibfield  {author} {\bibinfo {author} {\bibfnamefont {I.}~\bibnamefont
  {de~Vega}}\ and\ \bibinfo {author} {\bibfnamefont {D.}~\bibnamefont
  {Alonso}},\ }\bibfield  {title} {\bibinfo {title} {Dynamics of non-markovian
  open quantum systems},\ }\href {https://doi.org/10.1103/RevModPhys.89.015001}
  {\bibfield  {journal} {\bibinfo  {journal} {Rev. Mod. Phys.}\ }\textbf
  {\bibinfo {volume} {89}},\ \bibinfo {pages} {015001} (\bibinfo {year}
  {2017})}\BibitemShut {NoStop}%
\bibitem [{\citenamefont {Weimer}\ \emph {et~al.}(2021)\citenamefont {Weimer},
  \citenamefont {Kshetrimayum},\ and\ \citenamefont
  {Or\'us}}]{RevModPhys.93.015008}%
  \BibitemOpen
  \bibfield  {author} {\bibinfo {author} {\bibfnamefont {H.}~\bibnamefont
  {Weimer}}, \bibinfo {author} {\bibfnamefont {A.}~\bibnamefont
  {Kshetrimayum}},\ and\ \bibinfo {author} {\bibfnamefont {R.}~\bibnamefont
  {Or\'us}},\ }\bibfield  {title} {\bibinfo {title} {Simulation methods for
  open quantum many-body systems},\ }\href
  {https://doi.org/10.1103/RevModPhys.93.015008} {\bibfield  {journal}
  {\bibinfo  {journal} {Rev. Mod. Phys.}\ }\textbf {\bibinfo {volume} {93}},\
  \bibinfo {pages} {015008} (\bibinfo {year} {2021})}\BibitemShut {NoStop}%
\bibitem [{\citenamefont {Wilczek}(2012)}]{wilczek2012quantum_prl}%
  \BibitemOpen
  \bibfield  {author} {\bibinfo {author} {\bibfnamefont {F.}~\bibnamefont
  {Wilczek}},\ }\bibfield  {title} {\bibinfo {title} {Quantum time crystals},\
  }\href {https://doi.org/10.1103/PhysRevLett.109.160401} {\bibfield  {journal}
  {\bibinfo  {journal} {Phys. Rev. Lett.}\ }\textbf {\bibinfo {volume} {109}},\
  \bibinfo {pages} {160401} (\bibinfo {year} {2012})}\BibitemShut {NoStop}%
\bibitem [{\citenamefont {Zhang}\ \emph {et~al.}(2017)\citenamefont {Zhang},
  \citenamefont {Hess}, \citenamefont {Kyprianidis}, \citenamefont {Becker},
  \citenamefont {Lee}, \citenamefont {Smith}, \citenamefont {Pagano},
  \citenamefont {Potirniche}, \citenamefont {Potter}, \citenamefont
  {Vishwanath}, \citenamefont {Yao},\ and\ \citenamefont
  {Monroe}}]{zhang2017observation_nature}%
  \BibitemOpen
  \bibfield  {author} {\bibinfo {author} {\bibfnamefont {J.}~\bibnamefont
  {Zhang}}, \bibinfo {author} {\bibfnamefont {P.~W.}\ \bibnamefont {Hess}},
  \bibinfo {author} {\bibfnamefont {A.}~\bibnamefont {Kyprianidis}}, \bibinfo
  {author} {\bibfnamefont {P.}~\bibnamefont {Becker}}, \bibinfo {author}
  {\bibfnamefont {A.}~\bibnamefont {Lee}}, \bibinfo {author} {\bibfnamefont
  {J.}~\bibnamefont {Smith}}, \bibinfo {author} {\bibfnamefont
  {G.}~\bibnamefont {Pagano}}, \bibinfo {author} {\bibfnamefont {I.-D.}\
  \bibnamefont {Potirniche}}, \bibinfo {author} {\bibfnamefont {A.~C.}\
  \bibnamefont {Potter}}, \bibinfo {author} {\bibfnamefont {A.}~\bibnamefont
  {Vishwanath}}, \bibinfo {author} {\bibfnamefont {N.~Y.}\ \bibnamefont
  {Yao}},\ and\ \bibinfo {author} {\bibfnamefont {C.}~\bibnamefont {Monroe}},\
  }\bibfield  {title} {\bibinfo {title} {Observation of a discrete time
  crystal},\ }\href {https://doi.org/10.1038/nature21413} {\bibfield  {journal}
  {\bibinfo  {journal} {Nature}\ }\textbf {\bibinfo {volume} {543}},\ \bibinfo
  {pages} {217} (\bibinfo {year} {2017})}\BibitemShut {NoStop}%
\bibitem [{\citenamefont {Yao}\ \emph {et~al.}(2017)\citenamefont {Yao},
  \citenamefont {Potter}, \citenamefont {Potirniche},\ and\ \citenamefont
  {Vishwanath}}]{yao2017discrete_prl}%
  \BibitemOpen
  \bibfield  {author} {\bibinfo {author} {\bibfnamefont {N.~Y.}\ \bibnamefont
  {Yao}}, \bibinfo {author} {\bibfnamefont {A.~C.}\ \bibnamefont {Potter}},
  \bibinfo {author} {\bibfnamefont {I.-D.}\ \bibnamefont {Potirniche}},\ and\
  \bibinfo {author} {\bibfnamefont {A.}~\bibnamefont {Vishwanath}},\ }\bibfield
   {title} {\bibinfo {title} {Discrete time crystals: Rigidity, criticality,
  and realizations},\ }\href {https://doi.org/10.1103/PhysRevLett.118.030401}
  {\bibfield  {journal} {\bibinfo  {journal} {Phys. Rev. Lett.}\ }\textbf
  {\bibinfo {volume} {118}},\ \bibinfo {pages} {030401} (\bibinfo {year}
  {2017})}\BibitemShut {NoStop}%
\bibitem [{\citenamefont {Zaletel}\ \emph {et~al.}(2023)\citenamefont
  {Zaletel}, \citenamefont {Lukin}, \citenamefont {Monroe}, \citenamefont
  {Nayak}, \citenamefont {Wilczek},\ and\ \citenamefont
  {Yao}}]{RevModPhys.95.031001}%
  \BibitemOpen
  \bibfield  {author} {\bibinfo {author} {\bibfnamefont {M.~P.}\ \bibnamefont
  {Zaletel}}, \bibinfo {author} {\bibfnamefont {M.}~\bibnamefont {Lukin}},
  \bibinfo {author} {\bibfnamefont {C.}~\bibnamefont {Monroe}}, \bibinfo
  {author} {\bibfnamefont {C.}~\bibnamefont {Nayak}}, \bibinfo {author}
  {\bibfnamefont {F.}~\bibnamefont {Wilczek}},\ and\ \bibinfo {author}
  {\bibfnamefont {N.~Y.}\ \bibnamefont {Yao}},\ }\bibfield  {title} {\bibinfo
  {title} {Colloquium: Quantum and classical discrete time crystals},\ }\href
  {https://doi.org/10.1103/RevModPhys.95.031001} {\bibfield  {journal}
  {\bibinfo  {journal} {Rev. Mod. Phys.}\ }\textbf {\bibinfo {volume} {95}},\
  \bibinfo {pages} {031001} (\bibinfo {year} {2023})}\BibitemShut {NoStop}%
\bibitem [{\citenamefont {Chen}\ and\ \citenamefont
  {Zhang}(2023)}]{Chen2023_nc}%
  \BibitemOpen
  \bibfield  {author} {\bibinfo {author} {\bibfnamefont {Y.-H.}\ \bibnamefont
  {Chen}}\ and\ \bibinfo {author} {\bibfnamefont {X.}~\bibnamefont {Zhang}},\
  }\bibfield  {title} {\bibinfo {title} {Realization of an inherent time
  crystal in a dissipative many-body system},\ }\href
  {https://doi.org/10.1038/s41467-023-41905-3} {\bibfield  {journal} {\bibinfo
  {journal} {Nat. Commun.}\ }\textbf {\bibinfo {volume} {14}},\ \bibinfo
  {pages} {6161} (\bibinfo {year} {2023})}\BibitemShut {NoStop}%
\bibitem [{\citenamefont {Abanin}\ \emph {et~al.}(2019)\citenamefont {Abanin},
  \citenamefont {Altman}, \citenamefont {Bloch},\ and\ \citenamefont
  {Serbyn}}]{RevModPhys.91.021001}%
  \BibitemOpen
  \bibfield  {author} {\bibinfo {author} {\bibfnamefont {D.~A.}\ \bibnamefont
  {Abanin}}, \bibinfo {author} {\bibfnamefont {E.}~\bibnamefont {Altman}},
  \bibinfo {author} {\bibfnamefont {I.}~\bibnamefont {Bloch}},\ and\ \bibinfo
  {author} {\bibfnamefont {M.}~\bibnamefont {Serbyn}},\ }\bibfield  {title}
  {\bibinfo {title} {Colloquium: Many-body localization, thermalization, and
  entanglement},\ }\href {https://doi.org/10.1103/RevModPhys.91.021001}
  {\bibfield  {journal} {\bibinfo  {journal} {Rev. Mod. Phys.}\ }\textbf
  {\bibinfo {volume} {91}},\ \bibinfo {pages} {021001} (\bibinfo {year}
  {2019})}\BibitemShut {NoStop}%
\bibitem [{\citenamefont {Bloch}\ \emph {et~al.}(2008)\citenamefont {Bloch},
  \citenamefont {Dalibard},\ and\ \citenamefont {Zwerger}}]{RevModPhys.80.885}%
  \BibitemOpen
  \bibfield  {author} {\bibinfo {author} {\bibfnamefont {I.}~\bibnamefont
  {Bloch}}, \bibinfo {author} {\bibfnamefont {J.}~\bibnamefont {Dalibard}},\
  and\ \bibinfo {author} {\bibfnamefont {W.}~\bibnamefont {Zwerger}},\
  }\bibfield  {title} {\bibinfo {title} {Many-body physics with ultracold
  gases},\ }\href {https://doi.org/10.1103/RevModPhys.80.885} {\bibfield
  {journal} {\bibinfo  {journal} {Rev. Mod. Phys.}\ }\textbf {\bibinfo {volume}
  {80}},\ \bibinfo {pages} {885} (\bibinfo {year} {2008})}\BibitemShut
  {NoStop}%
\bibitem [{\citenamefont {Guo}\ \emph {et~al.}(2021)\citenamefont {Guo},
  \citenamefont {Cheng}, \citenamefont {Sun}, \citenamefont {Song},
  \citenamefont {Li}, \citenamefont {Wang}, \citenamefont {Ren}, \citenamefont
  {Dong}, \citenamefont {Zheng}, \citenamefont {Zhang}, \citenamefont
  {Mondaini}, \citenamefont {Fan},\ and\ \citenamefont {Wang}}]{Guo2021_np}%
  \BibitemOpen
  \bibfield  {author} {\bibinfo {author} {\bibfnamefont {Q.}~\bibnamefont
  {Guo}}, \bibinfo {author} {\bibfnamefont {C.}~\bibnamefont {Cheng}}, \bibinfo
  {author} {\bibfnamefont {Z.-H.}\ \bibnamefont {Sun}}, \bibinfo {author}
  {\bibfnamefont {Z.}~\bibnamefont {Song}}, \bibinfo {author} {\bibfnamefont
  {H.}~\bibnamefont {Li}}, \bibinfo {author} {\bibfnamefont {Z.}~\bibnamefont
  {Wang}}, \bibinfo {author} {\bibfnamefont {W.}~\bibnamefont {Ren}}, \bibinfo
  {author} {\bibfnamefont {H.}~\bibnamefont {Dong}}, \bibinfo {author}
  {\bibfnamefont {D.}~\bibnamefont {Zheng}}, \bibinfo {author} {\bibfnamefont
  {Y.-R.}\ \bibnamefont {Zhang}}, \bibinfo {author} {\bibfnamefont
  {R.}~\bibnamefont {Mondaini}}, \bibinfo {author} {\bibfnamefont
  {H.}~\bibnamefont {Fan}},\ and\ \bibinfo {author} {\bibfnamefont
  {H.}~\bibnamefont {Wang}},\ }\bibfield  {title} {\bibinfo {title}
  {Observation of energy-resolved many-body localization},\ }\href
  {https://doi.org/10.1038/s41567-020-1035-1} {\bibfield  {journal} {\bibinfo
  {journal} {Nat. Phys.}\ }\textbf {\bibinfo {volume} {17}},\ \bibinfo {pages}
  {234} (\bibinfo {year} {2021})}\BibitemShut {NoStop}%
\bibitem [{\citenamefont {Ueda}(2020)}]{ueda2020quantum_nrp}%
  \BibitemOpen
  \bibfield  {author} {\bibinfo {author} {\bibfnamefont {M.}~\bibnamefont
  {Ueda}},\ }\bibfield  {title} {\bibinfo {title} {Quantum equilibration,
  thermalization and prethermalization in ultracold atoms},\ }\href
  {https://doi.org/10.1038/s42254-020-0237-x} {\bibfield  {journal} {\bibinfo
  {journal} {Nat. Rev. Phys.}\ }\textbf {\bibinfo {volume} {2}},\ \bibinfo
  {pages} {669} (\bibinfo {year} {2020})}\BibitemShut {NoStop}%
\bibitem [{\citenamefont {Le}\ \emph {et~al.}(2023)\citenamefont {Le},
  \citenamefont {Zhang}, \citenamefont {Gopalakrishnan}, \citenamefont
  {Rigol},\ and\ \citenamefont {Weiss}}]{le2023observation_nature}%
  \BibitemOpen
  \bibfield  {author} {\bibinfo {author} {\bibfnamefont {Y.}~\bibnamefont
  {Le}}, \bibinfo {author} {\bibfnamefont {Y.}~\bibnamefont {Zhang}}, \bibinfo
  {author} {\bibfnamefont {S.}~\bibnamefont {Gopalakrishnan}}, \bibinfo
  {author} {\bibfnamefont {M.}~\bibnamefont {Rigol}},\ and\ \bibinfo {author}
  {\bibfnamefont {D.~S.}\ \bibnamefont {Weiss}},\ }\bibfield  {title} {\bibinfo
  {title} {Observation of hydrodynamization and local prethermalization in 1d
  bose gases},\ }\href {https://doi.org/10.1038/s41586-023-05979-9} {\bibfield
  {journal} {\bibinfo  {journal} {Nature}\ }\textbf {\bibinfo {volume} {618}},\
  \bibinfo {pages} {494} (\bibinfo {year} {2023})}\BibitemShut {NoStop}%
\bibitem [{\citenamefont {Gong}\ \emph {et~al.}(2018)\citenamefont {Gong},
  \citenamefont {Ashida}, \citenamefont {Kawabata}, \citenamefont {Takasan},
  \citenamefont {Higashikawa},\ and\ \citenamefont
  {Ueda}}]{Gong2018Topological_prx}%
  \BibitemOpen
  \bibfield  {author} {\bibinfo {author} {\bibfnamefont {Z.}~\bibnamefont
  {Gong}}, \bibinfo {author} {\bibfnamefont {Y.}~\bibnamefont {Ashida}},
  \bibinfo {author} {\bibfnamefont {K.}~\bibnamefont {Kawabata}}, \bibinfo
  {author} {\bibfnamefont {K.}~\bibnamefont {Takasan}}, \bibinfo {author}
  {\bibfnamefont {S.}~\bibnamefont {Higashikawa}},\ and\ \bibinfo {author}
  {\bibfnamefont {M.}~\bibnamefont {Ueda}},\ }\bibfield  {title} {\bibinfo
  {title} {Topological phases of non-hermitian systems},\ }\href
  {https://doi.org/10.1103/PhysRevX.8.031079} {\bibfield  {journal} {\bibinfo
  {journal} {Phys. Rev. X}\ }\textbf {\bibinfo {volume} {8}},\ \bibinfo {pages}
  {031079} (\bibinfo {year} {2018})}\BibitemShut {NoStop}%
\bibitem [{\citenamefont {Carleo}\ \emph {et~al.}(2019)\citenamefont {Carleo},
  \citenamefont {Cirac}, \citenamefont {Cranmer}, \citenamefont {Daudet},
  \citenamefont {Schuld}, \citenamefont {Tishby}, \citenamefont
  {Vogt-Maranto},\ and\ \citenamefont {Zdeborov\'a}}]{RevModPhys.91.045002}%
  \BibitemOpen
  \bibfield  {author} {\bibinfo {author} {\bibfnamefont {G.}~\bibnamefont
  {Carleo}}, \bibinfo {author} {\bibfnamefont {I.}~\bibnamefont {Cirac}},
  \bibinfo {author} {\bibfnamefont {K.}~\bibnamefont {Cranmer}}, \bibinfo
  {author} {\bibfnamefont {L.}~\bibnamefont {Daudet}}, \bibinfo {author}
  {\bibfnamefont {M.}~\bibnamefont {Schuld}}, \bibinfo {author} {\bibfnamefont
  {N.}~\bibnamefont {Tishby}}, \bibinfo {author} {\bibfnamefont
  {L.}~\bibnamefont {Vogt-Maranto}},\ and\ \bibinfo {author} {\bibfnamefont
  {L.}~\bibnamefont {Zdeborov\'a}},\ }\bibfield  {title} {\bibinfo {title}
  {Machine learning and the physical sciences},\ }\href
  {https://doi.org/10.1103/RevModPhys.91.045002} {\bibfield  {journal}
  {\bibinfo  {journal} {Rev. Mod. Phys.}\ }\textbf {\bibinfo {volume} {91}},\
  \bibinfo {pages} {045002} (\bibinfo {year} {2019})}\BibitemShut {NoStop}%
\bibitem [{\citenamefont {Rem}\ \emph {et~al.}(2019)\citenamefont {Rem},
  \citenamefont {K{\"a}ming}, \citenamefont {Tarnowski}, \citenamefont
  {Asteria}, \citenamefont {Fl{\"a}schner}, \citenamefont {Becker},
  \citenamefont {Sengstock},\ and\ \citenamefont
  {Weitenberg}}]{rem2019identifying}%
  \BibitemOpen
  \bibfield  {author} {\bibinfo {author} {\bibfnamefont {B.~S.}\ \bibnamefont
  {Rem}}, \bibinfo {author} {\bibfnamefont {N.}~\bibnamefont {K{\"a}ming}},
  \bibinfo {author} {\bibfnamefont {M.}~\bibnamefont {Tarnowski}}, \bibinfo
  {author} {\bibfnamefont {L.}~\bibnamefont {Asteria}}, \bibinfo {author}
  {\bibfnamefont {N.}~\bibnamefont {Fl{\"a}schner}}, \bibinfo {author}
  {\bibfnamefont {C.}~\bibnamefont {Becker}}, \bibinfo {author} {\bibfnamefont
  {K.}~\bibnamefont {Sengstock}},\ and\ \bibinfo {author} {\bibfnamefont
  {C.}~\bibnamefont {Weitenberg}},\ }\bibfield  {title} {\bibinfo {title}
  {Identifying quantum phase transitions using artificial neural networks on
  experimental data},\ }\href {https://doi.org/10.1038/s41567-019-0554-0}
  {\bibfield  {journal} {\bibinfo  {journal} {Nature Physics}\ }\textbf
  {\bibinfo {volume} {15}},\ \bibinfo {pages} {917} (\bibinfo {year}
  {2019})}\BibitemShut {NoStop}%
\bibitem [{\citenamefont {K{\"a}ming}\ \emph {et~al.}(2021)\citenamefont
  {K{\"a}ming}, \citenamefont {Dawid}, \citenamefont {Kottmann}, \citenamefont
  {Lewenstein}, \citenamefont {Sengstock}, \citenamefont {Dauphin},\ and\
  \citenamefont {Weitenberg}}]{kaming2021unsupervised}%
  \BibitemOpen
  \bibfield  {author} {\bibinfo {author} {\bibfnamefont {N.}~\bibnamefont
  {K{\"a}ming}}, \bibinfo {author} {\bibfnamefont {A.}~\bibnamefont {Dawid}},
  \bibinfo {author} {\bibfnamefont {K.}~\bibnamefont {Kottmann}}, \bibinfo
  {author} {\bibfnamefont {M.}~\bibnamefont {Lewenstein}}, \bibinfo {author}
  {\bibfnamefont {K.}~\bibnamefont {Sengstock}}, \bibinfo {author}
  {\bibfnamefont {A.}~\bibnamefont {Dauphin}},\ and\ \bibinfo {author}
  {\bibfnamefont {C.}~\bibnamefont {Weitenberg}},\ }\bibfield  {title}
  {\bibinfo {title} {Unsupervised machine learning of topological phase
  transitions from experimental data},\ }\href
  {https://doi.org/10.1088/2632-2153/abffe7} {\bibfield  {journal} {\bibinfo
  {journal} {Machine Learning: Science and Technology}\ }\textbf {\bibinfo
  {volume} {2}},\ \bibinfo {pages} {035037} (\bibinfo {year}
  {2021})}\BibitemShut {NoStop}%
\bibitem [{\citenamefont {van Nieuwenburg}\ \emph {et~al.}(2017)\citenamefont
  {van Nieuwenburg}, \citenamefont {Liu},\ and\ \citenamefont
  {Huber}}]{van2017_np}%
  \BibitemOpen
  \bibfield  {author} {\bibinfo {author} {\bibfnamefont {E.~P.~L.}\
  \bibnamefont {van Nieuwenburg}}, \bibinfo {author} {\bibfnamefont {Y.-H.}\
  \bibnamefont {Liu}},\ and\ \bibinfo {author} {\bibfnamefont {S.~D.}\
  \bibnamefont {Huber}},\ }\bibfield  {title} {\bibinfo {title} {Learning phase
  transitions by confusion},\ }\href {https://doi.org/10.1038/nphys4037}
  {\bibfield  {journal} {\bibinfo  {journal} {Nat. Phys.}\ }\textbf {\bibinfo
  {volume} {13}},\ \bibinfo {pages} {435} (\bibinfo {year} {2017})}\BibitemShut
  {NoStop}%
\bibitem [{\citenamefont {Lu}\ \emph {et~al.}(2020)\citenamefont {Lu},
  \citenamefont {Kim},\ and\ \citenamefont
  {Solja{\v{c}}i{\'c}}}]{Lu2020Extracting_prx}%
  \BibitemOpen
  \bibfield  {author} {\bibinfo {author} {\bibfnamefont {P.~Y.}\ \bibnamefont
  {Lu}}, \bibinfo {author} {\bibfnamefont {S.}~\bibnamefont {Kim}},\ and\
  \bibinfo {author} {\bibfnamefont {M.}~\bibnamefont {Solja{\v{c}}i{\'c}}},\
  }\bibfield  {title} {\bibinfo {title} {Extracting interpretable physical
  parameters from spatiotemporal systems using unsupervised learning},\ }\href
  {https://doi.org/10.1103/PhysRevX.10.031056} {\bibfield  {journal} {\bibinfo
  {journal} {Phys. Rev. X}\ }\textbf {\bibinfo {volume} {10}},\ \bibinfo
  {pages} {031056} (\bibinfo {year} {2020})}\BibitemShut {NoStop}%
\bibitem [{\citenamefont {Wang}(2016)}]{WangLei_2016_prb}%
  \BibitemOpen
  \bibfield  {author} {\bibinfo {author} {\bibfnamefont {L.}~\bibnamefont
  {Wang}},\ }\bibfield  {title} {\bibinfo {title} {Discovering phase
  transitions with unsupervised learning},\ }\href
  {https://doi.org/https://doi.org/10.1103/PhysRevB.94.195105} {\bibfield
  {journal} {\bibinfo  {journal} {Phys. Rev. B}\ }\textbf {\bibinfo {volume}
  {94}},\ \bibinfo {pages} {195105} (\bibinfo {year} {2016})}\BibitemShut
  {NoStop}%
\bibitem [{\citenamefont {Hu}\ \emph {et~al.}(2017)\citenamefont {Hu},
  \citenamefont {Singh},\ and\ \citenamefont {Scalettar}}]{Hu2017_pre}%
  \BibitemOpen
  \bibfield  {author} {\bibinfo {author} {\bibfnamefont {W.}~\bibnamefont
  {Hu}}, \bibinfo {author} {\bibfnamefont {R.~R.~P.}\ \bibnamefont {Singh}},\
  and\ \bibinfo {author} {\bibfnamefont {R.~T.}\ \bibnamefont {Scalettar}},\
  }\bibfield  {title} {\bibinfo {title} {Discovering phases, phase transitions,
  and crossovers through unsupervised machine learning: A critical
  examination},\ }\href {https://doi.org/10.1103/PhysRevE.95.062122} {\bibfield
   {journal} {\bibinfo  {journal} {Phys. Rev. E}\ }\textbf {\bibinfo {volume}
  {95}},\ \bibinfo {pages} {062122} (\bibinfo {year} {2017})}\BibitemShut
  {NoStop}%
\bibitem [{\citenamefont {Wang}\ and\ \citenamefont
  {Zhai}(2017)}]{Wang2017Machine_prb}%
  \BibitemOpen
  \bibfield  {author} {\bibinfo {author} {\bibfnamefont {C.}~\bibnamefont
  {Wang}}\ and\ \bibinfo {author} {\bibfnamefont {H.}~\bibnamefont {Zhai}},\
  }\bibfield  {title} {\bibinfo {title} {Machine learning of frustrated
  classical spin models. i. principal component analysis},\ }\href
  {https://doi.org/10.1103/PhysRevB.96.144432} {\bibfield  {journal} {\bibinfo
  {journal} {Phys. Rev. B}\ }\textbf {\bibinfo {volume} {96}},\ \bibinfo
  {pages} {144432} (\bibinfo {year} {2017})}\BibitemShut {NoStop}%
\bibitem [{\citenamefont {Rodriguez-Nieva}\ and\ \citenamefont
  {Scheurer}(2019{\natexlab{a}})}]{Rodriguez-Nieva2019_np}%
  \BibitemOpen
  \bibfield  {author} {\bibinfo {author} {\bibfnamefont {J.~F.}\ \bibnamefont
  {Rodriguez-Nieva}}\ and\ \bibinfo {author} {\bibfnamefont {M.~S.}\
  \bibnamefont {Scheurer}},\ }\bibfield  {title} {\bibinfo {title} {Identifying
  topological order through unsupervised machine learning},\ }\href
  {https://doi.org/10.1038/s41567-019-0512-x} {\bibfield  {journal} {\bibinfo
  {journal} {Nat. Phys.}\ }\textbf {\bibinfo {volume} {15}},\ \bibinfo {pages}
  {790} (\bibinfo {year} {2019}{\natexlab{a}})}\BibitemShut {NoStop}%
\bibitem [{\citenamefont {Yang}\ \emph {et~al.}(2020)\citenamefont {Yang},
  \citenamefont {Ren},\ and\ \citenamefont {Chen}}]{YangLong2020_prl}%
  \BibitemOpen
  \bibfield  {author} {\bibinfo {author} {\bibfnamefont {L.}~\bibnamefont
  {Yang}}, \bibinfo {author} {\bibfnamefont {J.}~\bibnamefont {Ren}},\ and\
  \bibinfo {author} {\bibfnamefont {H.}~\bibnamefont {Chen}},\ }\bibfield
  {title} {\bibinfo {title} {Unsupervised manifold clustering of topological
  phononics},\ }\href
  {https://doi.org/https://doi.org/10.1103/PhysRevLett.124.185501} {\bibfield
  {journal} {\bibinfo  {journal} {Phys. Rev. Lett.}\ }\textbf {\bibinfo
  {volume} {124}},\ \bibinfo {pages} {165502} (\bibinfo {year}
  {2020})}\BibitemShut {NoStop}%
\bibitem [{\citenamefont {Che}\ \emph {et~al.}(2020)\citenamefont {Che},
  \citenamefont {Gneiting}, \citenamefont {Liu},\ and\ \citenamefont
  {Nori}}]{Che2020Topological_prb}%
  \BibitemOpen
  \bibfield  {author} {\bibinfo {author} {\bibfnamefont {Y.}~\bibnamefont
  {Che}}, \bibinfo {author} {\bibfnamefont {C.}~\bibnamefont {Gneiting}},
  \bibinfo {author} {\bibfnamefont {T.}~\bibnamefont {Liu}},\ and\ \bibinfo
  {author} {\bibfnamefont {F.}~\bibnamefont {Nori}},\ }\bibfield  {title}
  {\bibinfo {title} {Topological quantum phase transitions retrieved through
  unsupervised machine learning},\ }\href
  {https://doi.org/10.1103/PhysRevB.102.134213} {\bibfield  {journal} {\bibinfo
   {journal} {Phys. Rev. B}\ }\textbf {\bibinfo {volume} {102}},\ \bibinfo
  {pages} {134213} (\bibinfo {year} {2020})}\BibitemShut {NoStop}%
\bibitem [{\citenamefont {Yu}\ and\ \citenamefont
  {Deng}(2021)}]{Yu_LiWei_2021_prl}%
  \BibitemOpen
  \bibfield  {author} {\bibinfo {author} {\bibfnamefont {L.-W.}\ \bibnamefont
  {Yu}}\ and\ \bibinfo {author} {\bibfnamefont {D.-L.}\ \bibnamefont {Deng}},\
  }\bibfield  {title} {\bibinfo {title} {Unsupervised learning of non-hermitian
  topological phases},\ }\href {https://doi.org/10.1103/PhysRevLett.126.240402}
  {\bibfield  {journal} {\bibinfo  {journal} {Phys. Rev. Lett.}\ }\textbf
  {\bibinfo {volume} {126}},\ \bibinfo {pages} {240402} (\bibinfo {year}
  {2021})}\BibitemShut {NoStop}%
\bibitem [{SM()}]{SM}%
  \BibitemOpen
  \href@noop {} {}\bibinfo {note} {See the Supplemental Material for additional
  details, including: additional discussion supporting the schematic
  illustration and the temporal-fluctuation amplification mechanism (Sec.~S1);
  hyperparameter settings for the DTC and AA models (Sec.~S2);
  silhouette-coefficient analysis and clustering diagnostics (Sec.~S3); the
  procedure for extracting phase-transition points via silhouette landscapes
  and learning-by-confusion (Sec.~S4); additional model benchmarks, including
  the Quantum East model and the Feingold--Peres model (Secs.~S5--S6), as well
  as an independent transverse-field Ising validation (Sec.~S7); an operational
  $S$-saturation criterion for selecting the boosted power $\beta$ (Sec.~S8);
  an ablation demonstrating that the Euclidean component is indispensable to
  the TFCAD metric (Sec.~S9); feature/observable selection guidelines for
  dynamical classification (Sec.~S10); and robustness against sampling-interval
  variations $\delta_t$ (Sec.~S11).}\BibitemShut {Stop}%
\bibitem [{\citenamefont {Batista}\ \emph {et~al.}(2011)\citenamefont
  {Batista}, \citenamefont {Wang},\ and\ \citenamefont
  {Keogh}}]{Gustavo2011_book}%
  \BibitemOpen
  \bibfield  {author} {\bibinfo {author} {\bibfnamefont {G.~E. A. P.~A.}\
  \bibnamefont {Batista}}, \bibinfo {author} {\bibfnamefont {X.}~\bibnamefont
  {Wang}},\ and\ \bibinfo {author} {\bibfnamefont {E.~J.}\ \bibnamefont
  {Keogh}},\ }\bibfield  {title} {\bibinfo {title} {A complexity-invariant
  distance measure for time series},\ }\href
  {https://doi.org/10.1137/1.9781611972818.60} {\bibfield  {journal} {\bibinfo
  {journal} {Proc. SIAM Int. Conf. Data Mining}\ ,\ \bibinfo {pages} {699}}
  (\bibinfo {year} {2011})}\BibitemShut {NoStop}%
\bibitem [{\citenamefont {West}(2014)}]{RevModPhys.86.1169}%
  \BibitemOpen
  \bibfield  {author} {\bibinfo {author} {\bibfnamefont {B.~J.}\ \bibnamefont
  {West}},\ }\bibfield  {title} {\bibinfo {title} {Colloquium: Fractional
  calculus view of complexity: A tutorial},\ }\href
  {https://doi.org/10.1103/RevModPhys.86.1169} {\bibfield  {journal} {\bibinfo
  {journal} {Rev. Mod. Phys.}\ }\textbf {\bibinfo {volume} {86}},\ \bibinfo
  {pages} {1169} (\bibinfo {year} {2014})}\BibitemShut {NoStop}%
\bibitem [{\citenamefont {Durr}\ and\ \citenamefont
  {Chakravarty}(2019)}]{Durr2019_prb}%
  \BibitemOpen
  \bibfield  {author} {\bibinfo {author} {\bibfnamefont {S.}~\bibnamefont
  {Durr}}\ and\ \bibinfo {author} {\bibfnamefont {S.}~\bibnamefont
  {Chakravarty}},\ }\bibfield  {title} {\bibinfo {title} {Unsupervised learning
  eigenstate phases of matter},\ }\href
  {https://doi.org/https://doi.org/10.1103/PhysRevB.100.075102} {\bibfield
  {journal} {\bibinfo  {journal} {Phys. Rev. B}\ }\textbf {\bibinfo {volume}
  {100}},\ \bibinfo {pages} {075102} (\bibinfo {year} {2019})}\BibitemShut
  {NoStop}%
\bibitem [{\citenamefont {Rodriguez-Nieva}\ and\ \citenamefont
  {Scheurer}(2019{\natexlab{b}})}]{RodriguezNieva2019_np}%
  \BibitemOpen
  \bibfield  {author} {\bibinfo {author} {\bibfnamefont {J.~F.}\ \bibnamefont
  {Rodriguez-Nieva}}\ and\ \bibinfo {author} {\bibfnamefont {M.~S.}\
  \bibnamefont {Scheurer}},\ }\bibfield  {title} {\bibinfo {title} {Identifying
  topological order through unsupervised machine learning},\ }\href
  {https://doi.org/10.1038/s41567-019-0512-x} {\bibfield  {journal} {\bibinfo
  {journal} {Nat. Phys.}\ }\textbf {\bibinfo {volume} {15}},\ \bibinfo {pages}
  {790} (\bibinfo {year} {2019}{\natexlab{b}})}\BibitemShut {NoStop}%
\bibitem [{\citenamefont {Zvyagintseva}\ \emph {et~al.}(2021)\citenamefont
  {Zvyagintseva}, \citenamefont {Nikitin}, \citenamefont {Kuandykov},
  \citenamefont {Ulyantsev},\ and\ \citenamefont
  {Zaikin}}]{Zvyagintseva2021_commphys}%
  \BibitemOpen
  \bibfield  {author} {\bibinfo {author} {\bibfnamefont {D.}~\bibnamefont
  {Zvyagintseva}}, \bibinfo {author} {\bibfnamefont {A.}~\bibnamefont
  {Nikitin}}, \bibinfo {author} {\bibfnamefont {I.}~\bibnamefont {Kuandykov}},
  \bibinfo {author} {\bibfnamefont {V.}~\bibnamefont {Ulyantsev}},\ and\
  \bibinfo {author} {\bibfnamefont {A.}~\bibnamefont {Zaikin}},\ }\bibfield
  {title} {\bibinfo {title} {Machine learning of phase transitions in nonlinear
  dynamics},\ }\href
  {https://doi.org/https://doi.org/10.1038/s42005-021-00755-5} {\bibfield
  {journal} {\bibinfo  {journal} {Commun. Phys.}\ }\textbf {\bibinfo {volume}
  {4}},\ \bibinfo {pages} {232} (\bibinfo {year} {2021})}\BibitemShut {NoStop}%
\bibitem [{ij()}]{ij}%
  \BibitemOpen
  \href@noop {} {}\bibinfo {note} {For simplicity, we now use subscripts $ij$
  to represent data sets $\mathbf{D}_i ,\mathbf{D}_j$, for example, $M_{ij}$
  means $M(\mathbf{D}_i,\mathbf{D}_j)$ and $C_i$ means
  $C(\mathbf{D}_i)$.}\BibitemShut {Stop}%
\bibitem [{\citenamefont {Lidiak}\ and\ \citenamefont
  {Gong}(2020)}]{lidiak2020unsupervised}%
  \BibitemOpen
  \bibfield  {author} {\bibinfo {author} {\bibfnamefont {A.}~\bibnamefont
  {Lidiak}}\ and\ \bibinfo {author} {\bibfnamefont {Z.}~\bibnamefont {Gong}},\
  }\bibfield  {title} {\bibinfo {title} {Unsupervised machine learning of
  quantum phase transitions using diffusion maps},\ }\href
  {https://doi.org/10.1103/PhysRevLett.125.225701} {\bibfield  {journal}
  {\bibinfo  {journal} {Phys. Rev. Lett.}\ }\textbf {\bibinfo {volume} {125}},\
  \bibinfo {pages} {225701} (\bibinfo {year} {2020})}\BibitemShut {NoStop}%
\bibitem [{\citenamefont {Ponte}\ \emph {et~al.}(2015)\citenamefont {Ponte},
  \citenamefont {Papi\ifmmode~\acute{c}\else \'{c}\fi{}}, \citenamefont
  {Huveneers},\ and\ \citenamefont {Abanin}}]{ponte2015manybody_prl}%
  \BibitemOpen
  \bibfield  {author} {\bibinfo {author} {\bibfnamefont {P.}~\bibnamefont
  {Ponte}}, \bibinfo {author} {\bibfnamefont {Z.}~\bibnamefont
  {Papi\ifmmode~\acute{c}\else \'{c}\fi{}}}, \bibinfo {author} {\bibfnamefont
  {F.~m.~c.}\ \bibnamefont {Huveneers}},\ and\ \bibinfo {author} {\bibfnamefont
  {D.~A.}\ \bibnamefont {Abanin}},\ }\bibfield  {title} {\bibinfo {title}
  {Many-body localization in periodically driven systems},\ }\href
  {https://doi.org/10.1103/PhysRevLett.114.140401} {\bibfield  {journal}
  {\bibinfo  {journal} {Phys. Rev. Lett.}\ }\textbf {\bibinfo {volume} {114}},\
  \bibinfo {pages} {140401} (\bibinfo {year} {2015})}\BibitemShut {NoStop}%
\bibitem [{\citenamefont {Lazarides}\ \emph {et~al.}(2015)\citenamefont
  {Lazarides}, \citenamefont {Das},\ and\ \citenamefont
  {Moessner}}]{lazarides2015fate_prl}%
  \BibitemOpen
  \bibfield  {author} {\bibinfo {author} {\bibfnamefont {A.}~\bibnamefont
  {Lazarides}}, \bibinfo {author} {\bibfnamefont {A.}~\bibnamefont {Das}},\
  and\ \bibinfo {author} {\bibfnamefont {R.}~\bibnamefont {Moessner}},\
  }\bibfield  {title} {\bibinfo {title} {Fate of many-body localization under
  periodic driving},\ }\href {https://doi.org/10.1103/PhysRevLett.115.030402}
  {\bibfield  {journal} {\bibinfo  {journal} {Phys. Rev. Lett.}\ }\textbf
  {\bibinfo {volume} {115}},\ \bibinfo {pages} {030402} (\bibinfo {year}
  {2015})}\BibitemShut {NoStop}%
\bibitem [{\citenamefont {Khemani}\ \emph {et~al.}(2016)\citenamefont
  {Khemani}, \citenamefont {Lazarides}, \citenamefont {Moessner},\ and\
  \citenamefont {Sondhi}}]{khemani2016phase_prl}%
  \BibitemOpen
  \bibfield  {author} {\bibinfo {author} {\bibfnamefont {V.}~\bibnamefont
  {Khemani}}, \bibinfo {author} {\bibfnamefont {A.}~\bibnamefont {Lazarides}},
  \bibinfo {author} {\bibfnamefont {R.}~\bibnamefont {Moessner}},\ and\
  \bibinfo {author} {\bibfnamefont {S.}~\bibnamefont {Sondhi}},\ }\bibfield
  {title} {\bibinfo {title} {Phase structure of driven quantum systems},\
  }\href {https://doi.org/10.1103/PhysRevLett.116.250401} {\bibfield  {journal}
  {\bibinfo  {journal} {Phys. Rev. Lett.}\ }\textbf {\bibinfo {volume} {116}},\
  \bibinfo {pages} {250401} (\bibinfo {year} {2016})}\BibitemShut {NoStop}%
\bibitem [{\citenamefont {Zhang}\ \emph {et~al.}(2023)\citenamefont {Zhang},
  \citenamefont {Wu}, \citenamefont {Qiu}, \citenamefont {Nan},\ and\
  \citenamefont {Li}}]{xiaopeng2023subexponential_prb}%
  \BibitemOpen
  \bibfield  {author} {\bibinfo {author} {\bibfnamefont {W.}~\bibnamefont
  {Zhang}}, \bibinfo {author} {\bibfnamefont {Y.}~\bibnamefont {Wu}}, \bibinfo
  {author} {\bibfnamefont {X.}~\bibnamefont {Qiu}}, \bibinfo {author}
  {\bibfnamefont {J.}~\bibnamefont {Nan}},\ and\ \bibinfo {author}
  {\bibfnamefont {X.}~\bibnamefont {Li}},\ }\bibfield  {title} {\bibinfo
  {title} {Subexponential critical slowing-down at a floquet time-crystal phase
  transition},\ }\href {https://doi.org/10.1103/PhysRevB.108.014307} {\bibfield
   {journal} {\bibinfo  {journal} {Phys. Rev. B}\ }\textbf {\bibinfo {volume}
  {108}},\ \bibinfo {pages} {014307} (\bibinfo {year} {2023})}\BibitemShut
  {NoStop}%
\bibitem [{\citenamefont {Moessner}\ and\ \citenamefont
  {Sondhi}(2017)}]{Moessner2017_np}%
  \BibitemOpen
  \bibfield  {author} {\bibinfo {author} {\bibfnamefont {R.}~\bibnamefont
  {Moessner}}\ and\ \bibinfo {author} {\bibfnamefont {S.~L.}\ \bibnamefont
  {Sondhi}},\ }\bibfield  {title} {\bibinfo {title} {Equilibration and order in
  quantum floquet matter},\ }\href {https://doi.org/10.1038/nphys4106}
  {\bibfield  {journal} {\bibinfo  {journal} {Nat. Phys.}\ }\textbf {\bibinfo
  {volume} {13}},\ \bibinfo {pages} {424} (\bibinfo {year} {2017})}\BibitemShut
  {NoStop}%
\bibitem [{\citenamefont {Quan}\ \emph {et~al.}(2006)\citenamefont {Quan},
  \citenamefont {Song}, \citenamefont {Liu}, \citenamefont {Zanardi},\ and\
  \citenamefont {Sun}}]{Quan_H_T_2006_prl}%
  \BibitemOpen
  \bibfield  {author} {\bibinfo {author} {\bibfnamefont {H.~T.}\ \bibnamefont
  {Quan}}, \bibinfo {author} {\bibfnamefont {Z.}~\bibnamefont {Song}}, \bibinfo
  {author} {\bibfnamefont {X.~F.}\ \bibnamefont {Liu}}, \bibinfo {author}
  {\bibfnamefont {P.}~\bibnamefont {Zanardi}},\ and\ \bibinfo {author}
  {\bibfnamefont {C.~P.}\ \bibnamefont {Sun}},\ }\bibfield  {title} {\bibinfo
  {title} {Decay of loschmidt echo enhanced by quantum criticality},\ }\href
  {https://doi.org/https://doi.org/10.1103/PhysRevLett.96.140604} {\bibfield
  {journal} {\bibinfo  {journal} {Phys. Rev. Lett.}\ }\textbf {\bibinfo
  {volume} {96}},\ \bibinfo {pages} {140604} (\bibinfo {year}
  {2006})}\BibitemShut {NoStop}%
\bibitem [{\citenamefont {Gorin}\ \emph {et~al.}(2006)\citenamefont {Gorin},
  \citenamefont {Prosen}, \citenamefont {Seligman},\ and\ \citenamefont
  {{\v{Z}}nidari{\v{c}}}}]{GORIN2006_pr}%
  \BibitemOpen
  \bibfield  {author} {\bibinfo {author} {\bibfnamefont {T.}~\bibnamefont
  {Gorin}}, \bibinfo {author} {\bibfnamefont {T.}~\bibnamefont {Prosen}},
  \bibinfo {author} {\bibfnamefont {T.~H.}\ \bibnamefont {Seligman}},\ and\
  \bibinfo {author} {\bibfnamefont {M.}~\bibnamefont {{\v{Z}}nidari{\v{c}}}},\
  }\bibfield  {title} {\bibinfo {title} {Dynamics of loschmidt echoes and
  fidelity decay},\ }\href
  {https://doi.org/https://doi.org/10.1016/j.physrep.2006.09.003} {\bibfield
  {journal} {\bibinfo  {journal} {Phys. Rep.}\ }\textbf {\bibinfo {volume}
  {435}},\ \bibinfo {pages} {33} (\bibinfo {year} {2006})}\BibitemShut
  {NoStop}%
\bibitem [{\citenamefont {Yan}\ \emph {et~al.}(2020)\citenamefont {Yan},
  \citenamefont {Cincio},\ and\ \citenamefont
  {Zurek}}]{yan2020information_prl}%
  \BibitemOpen
  \bibfield  {author} {\bibinfo {author} {\bibfnamefont {B.}~\bibnamefont
  {Yan}}, \bibinfo {author} {\bibfnamefont {L.}~\bibnamefont {Cincio}},\ and\
  \bibinfo {author} {\bibfnamefont {W.~H.}\ \bibnamefont {Zurek}},\ }\bibfield
  {title} {\bibinfo {title} {Information scrambling and loschmidt echo},\
  }\href {https://doi.org/10.1103/PhysRevLett.124.160603} {\bibfield  {journal}
  {\bibinfo  {journal} {Phys. Rev. lett.}\ }\textbf {\bibinfo {volume} {124}},\
  \bibinfo {pages} {160603} (\bibinfo {year} {2020})}\BibitemShut {NoStop}%
\bibitem [{\citenamefont {Martinez}\ \emph {et~al.}(2016)\citenamefont
  {Martinez}, \citenamefont {Muschik}, \citenamefont {Schindler}, \citenamefont
  {Nigg}, \citenamefont {Erhard}, \citenamefont {Heyl}, \citenamefont {Hauke},
  \citenamefont {Dalmonte}, \citenamefont {Monz}, \citenamefont {Zoller},\ and\
  \citenamefont {Blatt}}]{martinez2016real}%
  \BibitemOpen
  \bibfield  {author} {\bibinfo {author} {\bibfnamefont {E.~A.}\ \bibnamefont
  {Martinez}}, \bibinfo {author} {\bibfnamefont {C.~A.}\ \bibnamefont
  {Muschik}}, \bibinfo {author} {\bibfnamefont {P.}~\bibnamefont {Schindler}},
  \bibinfo {author} {\bibfnamefont {D.}~\bibnamefont {Nigg}}, \bibinfo {author}
  {\bibfnamefont {A.}~\bibnamefont {Erhard}}, \bibinfo {author} {\bibfnamefont
  {M.}~\bibnamefont {Heyl}}, \bibinfo {author} {\bibfnamefont {P.}~\bibnamefont
  {Hauke}}, \bibinfo {author} {\bibfnamefont {M.}~\bibnamefont {Dalmonte}},
  \bibinfo {author} {\bibfnamefont {T.}~\bibnamefont {Monz}}, \bibinfo {author}
  {\bibfnamefont {P.}~\bibnamefont {Zoller}},\ and\ \bibinfo {author}
  {\bibfnamefont {R.}~\bibnamefont {Blatt}},\ }\bibfield  {title} {\bibinfo
  {title} {Real-time dynamics of lattice gauge theories with a few-qubit
  quantum computer},\ }\href {https://doi.org/10.1038/nature18318} {\bibfield
  {journal} {\bibinfo  {journal} {Nature}\ }\textbf {\bibinfo {volume} {534}},\
  \bibinfo {pages} {516} (\bibinfo {year} {2016})}\BibitemShut {NoStop}%
\bibitem [{\citenamefont {Jurcevic}\ \emph {et~al.}(2017)\citenamefont
  {Jurcevic}, \citenamefont {Shen}, \citenamefont {Hauke}, \citenamefont
  {Maier}, \citenamefont {Brydges}, \citenamefont {Hempel}, \citenamefont
  {Lanyon}, \citenamefont {Heyl}, \citenamefont {Blatt},\ and\ \citenamefont
  {Roos}}]{jurcevic2017direct}%
  \BibitemOpen
  \bibfield  {author} {\bibinfo {author} {\bibfnamefont {P.}~\bibnamefont
  {Jurcevic}}, \bibinfo {author} {\bibfnamefont {H.}~\bibnamefont {Shen}},
  \bibinfo {author} {\bibfnamefont {P.}~\bibnamefont {Hauke}}, \bibinfo
  {author} {\bibfnamefont {C.}~\bibnamefont {Maier}}, \bibinfo {author}
  {\bibfnamefont {T.}~\bibnamefont {Brydges}}, \bibinfo {author} {\bibfnamefont
  {C.}~\bibnamefont {Hempel}}, \bibinfo {author} {\bibfnamefont {B.~P.}\
  \bibnamefont {Lanyon}}, \bibinfo {author} {\bibfnamefont {M.}~\bibnamefont
  {Heyl}}, \bibinfo {author} {\bibfnamefont {R.}~\bibnamefont {Blatt}},\ and\
  \bibinfo {author} {\bibfnamefont {C.~F.}\ \bibnamefont {Roos}},\ }\bibfield
  {title} {\bibinfo {title} {Direct observation of dynamical quantum phase
  transitions in an interacting many-body system},\ }\href
  {https://doi.org/10.1103/PhysRevLett.119.080501} {\bibfield  {journal}
  {\bibinfo  {journal} {Phys. Rev. Lett.}\ }\textbf {\bibinfo {volume} {119}},\
  \bibinfo {pages} {080501} (\bibinfo {year} {2017})}\BibitemShut {NoStop}%
\bibitem [{\citenamefont {Singh}\ \emph {et~al.}(2019)\citenamefont {Singh},
  \citenamefont {Fujiwara}, \citenamefont {Geiger}, \citenamefont {Simmons},
  \citenamefont {Lipatov}, \citenamefont {Cao}, \citenamefont {Dotti},
  \citenamefont {Rajagopal}, \citenamefont {Senaratne}, \citenamefont
  {Shimasaki}, \citenamefont {Heyl}, \citenamefont {Eckardt},\ and\
  \citenamefont {Weld}}]{singh2019quantifying}%
  \BibitemOpen
  \bibfield  {author} {\bibinfo {author} {\bibfnamefont {K.}~\bibnamefont
  {Singh}}, \bibinfo {author} {\bibfnamefont {C.~J.}\ \bibnamefont {Fujiwara}},
  \bibinfo {author} {\bibfnamefont {Z.~A.}\ \bibnamefont {Geiger}}, \bibinfo
  {author} {\bibfnamefont {E.~Q.}\ \bibnamefont {Simmons}}, \bibinfo {author}
  {\bibfnamefont {M.}~\bibnamefont {Lipatov}}, \bibinfo {author} {\bibfnamefont
  {A.}~\bibnamefont {Cao}}, \bibinfo {author} {\bibfnamefont {P.}~\bibnamefont
  {Dotti}}, \bibinfo {author} {\bibfnamefont {S.~V.}\ \bibnamefont
  {Rajagopal}}, \bibinfo {author} {\bibfnamefont {R.}~\bibnamefont
  {Senaratne}}, \bibinfo {author} {\bibfnamefont {T.}~\bibnamefont
  {Shimasaki}}, \bibinfo {author} {\bibfnamefont {M.}~\bibnamefont {Heyl}},
  \bibinfo {author} {\bibfnamefont {A.}~\bibnamefont {Eckardt}},\ and\ \bibinfo
  {author} {\bibfnamefont {D.~M.}\ \bibnamefont {Weld}},\ }\bibfield  {title}
  {\bibinfo {title} {Quantifying and controlling prethermal nonergodicity in
  interacting floquet matter},\ }\href
  {https://doi.org/10.1103/PhysRevX.9.041021} {\bibfield  {journal} {\bibinfo
  {journal} {Phys. Rev. X}\ }\textbf {\bibinfo {volume} {9}},\ \bibinfo {pages}
  {041021} (\bibinfo {year} {2019})}\BibitemShut {NoStop}%
\bibitem [{\citenamefont {Karch}\ \emph {et~al.}()\citenamefont {Karch},
  \citenamefont {Bandyopadhyay}, \citenamefont {Sun}, \citenamefont {Impertro},
  \citenamefont {Huh}, \citenamefont {Rodríguez}, \citenamefont {Wienand},
  \citenamefont {Ketterle}, \citenamefont {Heyl}, \citenamefont {Polkovnikov},
  \citenamefont {Bloch},\ and\ \citenamefont
  {Aidelsburger}}]{karch2025probing}%
  \BibitemOpen
  \bibfield  {author} {\bibinfo {author} {\bibfnamefont {S.}~\bibnamefont
  {Karch}}, \bibinfo {author} {\bibfnamefont {S.}~\bibnamefont
  {Bandyopadhyay}}, \bibinfo {author} {\bibfnamefont {Z.-H.}\ \bibnamefont
  {Sun}}, \bibinfo {author} {\bibfnamefont {A.}~\bibnamefont {Impertro}},
  \bibinfo {author} {\bibfnamefont {S.}~\bibnamefont {Huh}}, \bibinfo {author}
  {\bibfnamefont {I.~P.}\ \bibnamefont {Rodríguez}}, \bibinfo {author}
  {\bibfnamefont {J.~F.}\ \bibnamefont {Wienand}}, \bibinfo {author}
  {\bibfnamefont {W.}~\bibnamefont {Ketterle}}, \bibinfo {author}
  {\bibfnamefont {M.}~\bibnamefont {Heyl}}, \bibinfo {author} {\bibfnamefont
  {A.}~\bibnamefont {Polkovnikov}}, \bibinfo {author} {\bibfnamefont
  {I.}~\bibnamefont {Bloch}},\ and\ \bibinfo {author} {\bibfnamefont
  {M.}~\bibnamefont {Aidelsburger}},\ }\bibfield  {title} {\bibinfo {title}
  {Probing quantum many-body dynamics using subsystem {Loschmidt} echos},\
  }\href@noop {} {\ }\bibinfo {note}
  {\href{https://arxiv.org/abs/2501.16995}{arXiv:2501.16995(2025)}}\BibitemShut
  {NoStop}%
\bibitem [{\citenamefont {Aubry}\ and\ \citenamefont
  {Andr\'{e}'}(1980)}]{Aubry_1980}%
  \BibitemOpen
  \bibfield  {author} {\bibinfo {author} {\bibfnamefont {S.}~\bibnamefont
  {Aubry}}\ and\ \bibinfo {author} {\bibfnamefont {G.}~\bibnamefont
  {Andr\'{e}'}},\ }\bibfield  {title} {\bibinfo {title} {Localization in
  one-dimensional lattice structures},\ }\href@noop {} {\bibfield  {journal}
  {\bibinfo  {journal} {Ann. Isr. Phys. Soc.}\ }\textbf {\bibinfo {volume}
  {3}},\ \bibinfo {pages} {133} (\bibinfo {year} {1980})}\BibitemShut {NoStop}%
\bibitem [{\citenamefont {Iyer}\ \emph {et~al.}(2013)\citenamefont {Iyer},
  \citenamefont {Oganesyan}, \citenamefont {Refael},\ and\ \citenamefont
  {Huse}}]{Iyer2013_prb}%
  \BibitemOpen
  \bibfield  {author} {\bibinfo {author} {\bibfnamefont {S.}~\bibnamefont
  {Iyer}}, \bibinfo {author} {\bibfnamefont {V.}~\bibnamefont {Oganesyan}},
  \bibinfo {author} {\bibfnamefont {G.}~\bibnamefont {Refael}},\ and\ \bibinfo
  {author} {\bibfnamefont {D.~A.}\ \bibnamefont {Huse}},\ }\bibfield  {title}
  {\bibinfo {title} {Many-body localization in a quasiperiodic system},\ }\href
  {https://doi.org/10.1103/PhysRevB.87.134202} {\bibfield  {journal} {\bibinfo
  {journal} {Phys. Rev. B}\ }\textbf {\bibinfo {volume} {87}},\ \bibinfo
  {pages} {134202} (\bibinfo {year} {2013})}\BibitemShut {NoStop}%
\bibitem [{\citenamefont {Xu}\ \emph {et~al.}(2019)\citenamefont {Xu},
  \citenamefont {Li}, \citenamefont {Hsu}, \citenamefont {Swingle},\ and\
  \citenamefont {Das~Sarma}}]{xu2019butterfly_prr}%
  \BibitemOpen
  \bibfield  {author} {\bibinfo {author} {\bibfnamefont {S.}~\bibnamefont
  {Xu}}, \bibinfo {author} {\bibfnamefont {X.}~\bibnamefont {Li}}, \bibinfo
  {author} {\bibfnamefont {Y.-T.}\ \bibnamefont {Hsu}}, \bibinfo {author}
  {\bibfnamefont {B.}~\bibnamefont {Swingle}},\ and\ \bibinfo {author}
  {\bibfnamefont {S.}~\bibnamefont {Das~Sarma}},\ }\bibfield  {title} {\bibinfo
  {title} {Butterfly effect in interacting aubry-andre model: Thermalization,
  slow scrambling, and many-body localization},\ }\href
  {https://doi.org/10.1103/PhysRevResearch.1.032039} {\bibfield  {journal}
  {\bibinfo  {journal} {Phys. Rev. Res.}\ }\textbf {\bibinfo {volume} {1}},\
  \bibinfo {pages} {032039} (\bibinfo {year} {2019})}\BibitemShut {NoStop}%
\end{thebibliography}%

\end{document}